\documentclass[12pt]{article}
\usepackage{epsfig}
\catcode`@=11
\newcount\@tempcntc
\def\@citex[#1]#2{\if@filesw\immediate\write\@auxout{\string\citation{#2}}\fi
  \@tempcnta\z@\@tempcntb\m@ne\def\@citea{}\@cite{\@for\@citeb:=#2\do
    {\@ifundefined
       {b@\@citeb}{\@citeo\@tempcntb\m@ne\@citea\def\@citea{,}{\bf ?}\@warning
       {Citation `\@citeb' on page \thepage \space undefined}}%
    {\setbox\z@\hbox{\global\@tempcntc0\csname b@\@citeb\endcsname\relax}%
     \ifnum\@tempcntc=\z@ \@citeo\@tempcntb\m@ne
       \@citea\def\@citea{,}\hbox{\csname b@\@citeb\endcsname}%
     \else
      \advance\@tempcntb\@ne
      \ifnum\@tempcntb=\@tempcntc
      \else\advance\@tempcntb\m@ne\@citeo
      \@tempcnta\@tempcntc\@tempcntb\@tempcntc\fi\fi}}\@citeo}{#1}}
\def\@citeo{\ifnum\@tempcnta>\@tempcntb\else\@citea\def\@citea{,}%
  \ifnum\@tempcnta=\@tempcntb\the\@tempcnta\else
   {\advance\@tempcnta\@ne\ifnum\@tempcnta=\@tempcntb \else \def\@citea{--}\fi
    \advance\@tempcnta\m@ne\the\@tempcnta\@citea\the\@tempcntb}\fi\fi}
\catcode`@=12
\def\real{\mathop{\rm Re}\nolimits}
\def\imag{\mathop{\rm Im}\nolimits}
\def\slp{p\kern-5pt/}
\def\sls{s\kern-5pt/}
\def\as{\alpha_s}
\def\CF{C_{\scriptscriptstyle F}}
\def\NC{N_{\scriptscriptstyle C}}
\def\Li{\mbox{$\mbox{\rm Li}_2$}}
\def\Frac#1#2{\mbox{$\textstyle{#1\over#2}$}}
\def\eps{\varepsilon}
\def\ct{\cos\theta}
\def\cz{\chi_{\scriptscriptstyle Z}}
\def\MZ{M_{\scriptscriptstyle Z}}

\def\Pl{P^\ell}
\def\sxi{\mbox{$\sqrt\xi$}}
\def\jint{\int{dy\,dz\over(1-y)^2-\xi}\:\:}
\begin{document}
\thispagestyle{empty}
\begin{flushright}
MZ-TH/95-19 \\[-0.2cm]
FTUV/95-52 \\[-0.2cm]
IFIC/95-54 \\[-0.2cm]
hep-ph/9601313 \\[-0.2cm]
January 1996 \\[-0.2cm]
\end{flushright}
\begin{center}

{\Large\bf Polar Angle Dependence of the}\\[.3cm]
{\Large\bf Longitudinal Polarization of Quarks}\\[.3cm]
{\Large\bf Produced in \boldmath{$e^+e^-$}-Annihilation}\\[.75cm]

{\large S.~Groote$^*$, J.G.~K\"orner\footnote{Supported in part
by the BMFT, FRG, under contract 06MZ566,\\\hbox{\qquad}
and by HUCAM, EU, under contract CHRX-CT94-0579}} \\[.4cm]
Institut f\"ur Physik, Johannes Gutenberg-Universit\"at \\
Staudingerweg 7, D-55099 Mainz, Germany. \\[.5cm]

{\large M.M.~Tung\footnote[7]{Feodor-Lynen Fellow}} \\[.4cm]
Departament de F\'\i sica Te\`orica, Universitat~de Val\`encia \\
and IFIC, Centre Mixte Universitat Val\`encia --- CSIC, \\
C/ Dr.~Moliner, 50, E-46100 Burjassot (Val\`encia), Spain.
\end{center}
\vspace{.75cm}
\centerline {\bf ABSTRACT}
\noindent
We calculate one-loop radiative QCD corrections to the three polarized and 
unpolarized structure functions that determine the beam-quark polar angle 
dependence of the longitudinal polarization of light and heavy quarks 
produced in $e^+e^-$-annihilations. We present analytical and numerical 
results for the longitudinal polarization and its polar angle dependence. We 
discuss in some detail the zero-mass limit of our results and the role of 
the anomalous spin-flip contributions to the polarization observables in 
the zero-mass limit. Our discussion includes transverse and longitudinal 
beam polarization effects.

\newpage

\section{Introduction}
This is the fifth and final paper in a series of papers devoted to the 
$O(\as)$ determination of the polarization of quarks produced in 
$e^+e^-$-annihilations. In the first paper of this series~\cite{forward1} 
we calculated the mean longitudinal polarization of the quark,%
\footnote{In~\cite{forward3} we called the quark's longitudinal polarization 
its ``alignment'' polarization in order to set it apart from the term 
``longitudinal component'' of the exchanged gauge boson. In this paper we 
revert to the more accepted terminology ``longitudinal polarization''.} 
i.e.\ the longitudinal polarization averaged with respect to the relative 
beam-event orientation. One of us derived convenient Schwinger-type 
representations for the structure functions appearing in the polarization 
expressions in~\cite{forward2}. In a third paper we determined the 
contribution of the longitudinal component of the exchanged gauge boson to 
the quark's longitudinal polarization, which vanishes at the Born term 
level~\cite{forward3}. The fourth paper~\cite{forward4} gave results on the 
two transverse components of the quark's polarization (perpendicular and 
normal, also sometimes referred to as in-the-plane and out-of-the-plane 
transverse polarization). In the present final piece of work we present our 
results on the full polar angle dependence of the quark's longitudinal 
polarization with respect to the electron beam direction including beam 
polarization effects.

The full determination of the longitudinal polarization of the quark 
involves the calculation of three polarized and three unpolarized structure 
functions which are conveniently chosen as the helicity structure functions 
$H_{U,L,F}^\ell$ and $H_{U,L,F}$, respectively. The subscripts $U$, $L$ and 
$F$ label the relevant density matrix elements of the exchanged gauge 
boson ($\gamma_V$,$Z$), where the polar angle dependence of the 
contributions of the three structure functions is given by $(1+\cos^2\theta)$ 
for $U$, $\sin^2\theta$ for $L$ and $\cos\theta$ for $F$. We mention that 
some of the unpolarized helicity structure functions have been calculated 
before, i.e.\ the vector/vector ($VV$) contribution to $H_U$ and 
$H_L$~\cite{forward5}\footnote{The $VV$-contribution to $H_U+H_L$ 
has been given long time ago in the context of QED (see e.g.~\cite{forward6})} 
and the vector/axial-vector ($VA$) contribution to $H_F$~\cite{forward7}. 
We have verified the results of~\cite{forward5} and~\cite{forward7} and 
present new results on the axial-vector/axial-vector ($AA$) contribution to 
$H_U$ and $H_L$ in this paper, where the $AA$-contribution to 
$H_{U+L}=H_U+H_L$ was already written down in~\cite{forward1,forward2}. As 
concerns the polarized structure functions our results on $H_{U+L}^\ell$ 
and $H_L^\ell$ where given in~\cite{forward1,forward2,forward3}. This paper 
includes new results on the polarized structure functions $H_U^\ell$ and 
$H_F^\ell$ that are necessary to determine the full $\ct$-dependence of the 
quark's longitudinal polarization.

The paper is structured such that we start in Sec.~2 by listing the four 
independent tree-graph components of the hadron tensor ($VV$, $AA$, $V\!A$ 
and $AV$) including its polarization dependence. The relevant helicity 
components of the structure functions $H_{U,L,F}$ and $H_{U,L,F}^\ell$ are 
obtained by covariant projections. We then proceed to integrate the 
projected tree-graph contributions over the full three-body phase-space and 
give analytical results for the once- and twice-integrated tree-graph 
contributions. As is well familiar by now, the final expressions contain 
infrared (IR) divergences which we choose to regularize by introducing a 
small gluon mass. The tree-graph IR divergencies are cancelled by the 
corresponding IR divergencies of the one-loop contributions. Thus by adding 
in the loop contributions we finally arrive at finite results relevant for 
the total inclusive cross section integrated over the hard {\em and\/} the 
soft regions of the energy of the gluon as presented later on. In Sec.~2 we 
also detail the dependence of the polarization on the electroweak parameters 
including a discussion of beam polarization effects.

In Sec.~3 we focus on the soft-gluon region and present results on the 
polar angle dependence of the longitudinal polarization of the quark with 
their typical logarithmic dependence on the gluon-energy cut. The hard-gluon
contribution is given by the complement of the soft-gluon contribution, 
i.e.\ as the difference of the full $O(\as)$ result presented in Sec.~4 and 
the soft-gluon contribution. Since we provide numerically stable expressions 
for the latter two contributions, the hard-gluon contribution can be 
evaluated in a numerically stable way, if the need arises. In Sec.~5 we 
consider the zero quark mass case and calculate the three relevant polarized 
and unpolarized structure functions using helicity methods and dimensional 
reduction as regularization method. In this way one can keep out of the way 
of the notorious $\gamma_5$-problem when calculating e.g. the 
$V\!A$-contribution to the $F$-type unpolarized structure function. We 
compare the mass-zero results with the mass-zero limit of the corresponding 
structure function expressions in Sec.~4 and identify the global anomalous 
spin-flip contributions to the QCD$(m\rightarrow 0)$ polarized structure 
functions. Finally, Sec.~6 contains our summary and our conclusions. In 
Appendix~A we catalogue some integrals that appear in the tree-graph 
integrations in Sec.~2, in Appendix~B we list some standard $O(\as)$ rate 
functions needed for the cross section expressions in Sec.~4. In Appendix~C, 
finally, we consider the case of polarized and unpolarized quark production 
from transversely and longitudinally polarized $e^+e^-$-beams.

\begin{figure}\begin{center}
\epsfig{figure=born01.eps, scale=0.6}\qquad
\epsfig{figure=born02.eps, scale=0.6}\nonumber\\
\epsfig{figure=born03.eps, scale=0.6}\qquad
\epsfig{figure=born04.eps, scale=0.6}
\caption{\label{fig1}Born term and $O(\as)$ contributions to 
$e^+e^-\rightarrow q\bar q(g)$. One-loop wave function renormalization 
graphs are omitted}
\end{center}\end{figure}
\section{\boldmath{$O(\as)$} tree-graph contributions}
For the three body process
$(\gamma_V,Z)\rightarrow q(p_1)+\overline{q}(p_2)+g(p_3)$ (see
Fig.~\ref{fig1}) we define a polarized hadron tensor according to
($q=p_1+p_2+p_3$)
\begin{equation}\label{eqn1}
H_{\mu\nu}(q,p_1,p_2,s)=\sum_{\overline{q},g\mbox{\scriptsize\ spins}}
  \langle q\,\overline{q}\,g|j_\mu|0\rangle\langle 0|j_\nu^\dagger|
  q\,\overline{q}\,g\rangle.
\end{equation}
The hadron tensor depends on the vector ($V$) and axial-vector ($A$) 
composition of the currents in Eq.~(\ref{eqn1}). One has altogether four 
independent components $H_{\mu\nu}^i$ ($i=1,2,3,4$) which are defined 
according to ($V$: $\gamma_\mu$, $A$: $\gamma_\mu\gamma_5$)
\begin{eqnarray}
H_{\mu\nu}^1&=&\frac12(H_{\mu\nu}^{VV}+H_{\mu\nu}^{AA})\qquad
H_{\mu\nu}^2\ =\ \frac12(H_{\mu\nu}^{VV}-H_{\mu\nu}^{AA})\label{eqn2}\\
H_{\mu\nu}^3&=&\frac i2(H_{\mu\nu}^{V\!A}-H_{\mu\nu}^{AV})\qquad
H_{\mu\nu}^4\ =\ \frac12(H_{\mu\nu}^{V\!A}+H_{\mu\nu}^{AV}).\nonumber
\end{eqnarray}
For the tree-graph contribution one calculates
\begin{eqnarray}
H_{\mu\nu}^1&=&\frac{g_s^2\NC\CF}{2y^2z^2q^2}\Bigg[\ 
  q^2(-(2-\xi)(\xi(y+z)^2-4yz(1-y-z))+4yz(y^2+z^2))g_{\mu\nu}\nonumber\\&&
  +16y^2zp_{1\mu}p_{1\nu}+4(\xi y^2-4yz+2\xi yz+2y^2z+\xi z^2+2yz^2)
  (p_{1\mu}p_{2\nu}+p_{2\mu}p_{1\nu})\nonumber\\&&
  +16yz^2p_{2\mu}p_{2\nu}+4z(-2y+\xi y+2y^2+\xi z)
  (p_{1\mu}p_{3\nu}+p_{3\mu}p_{1\nu})\nonumber\\&&
  +4y(\xi y-2z+\xi z+2z^2)(p_{2\mu}p_{3\nu}+p_{3\mu}p_{2\nu})\ \Bigg]
  \nonumber\\&&
  +\frac{2img_s^2\NC\CF}{y^2z^2q^4}\Bigg[\ q^2(2-\xi)y^2\eps(\mu\nu p_1s)
  -q^2(4yz-2\xi yz+2y^2z-2yz^2)\eps(\mu\nu p_2s)\nonumber\\&&
  +q^2(2y^2-\xi y^2-2yz+2\xi yz+\xi z^2)\eps(\mu\nu p_3s)\nonumber\\&&
  -4y^2(p_{1\mu}+p_{3\mu})\eps(\nu p_1p_2s)
  +4y^2(p_{1\nu}+p_{3\nu})\eps(\mu p_1p_2s)\nonumber\\&&
  +4y(yp_{1\mu}+zp_{2\mu}+yp_{3\mu})\eps(\nu p_2p_3s)
  -4y(yp_{1\nu}+zp_{2\nu}+yp_{3\nu})\eps(\mu p_2p_3s)
  \ \Bigg]\\ \nonumber
  \\
H_{\mu\nu}^2&=&\frac{2m^2g_s^2\NC\CF}{y^2z^2q^4}\Bigg[ 
  -q^2(\xi(y+z)^2-4yz(1-y-z))g_{\mu\nu}+8yzp_{3\mu}p_{3\nu}\ \Bigg]
  \nonumber\\&& +\frac{2img_s^2\NC\CF}{y^2z^2q^4}\Bigg[\
  q^2(\xi y^2-2yz+\xi yz+\xi z^2+4yz^2)\eps(\mu\nu p_1s)\nonumber\\&&
  -q^2\xi z(y-z)\eps(\mu\nu p_2s)
  +q^2\xi y(y-z)\eps(\mu\nu p_3s)\nonumber\\&&
  -4yz(p_{1\mu}+p_{3\mu})\eps(\nu p_1p_2s)
  +4yz(p_{1\nu}+p_{3\nu})\eps(\mu p_1p_2s)\nonumber\\&&
  -4yz(p_{1\mu}-p_{2\mu}+p_{3\mu})\eps(\nu p_1p_3s)
  +4yz(p_{1\nu}-p_{2\nu}+p_{3\nu})\eps(\mu p_1p_3s)
  \ \Bigg]\\ \nonumber
  \\
H_{\mu\nu}^3&=&\frac{2img_s^2\NC\CF}{y^2z^2q^4}\Bigg[
  -4yz(p_3s)(p_{1\mu}p_{2\nu}-p_{2\mu}p_{1\nu})
  +4yz(p_2s+p_3s)(p_{1\mu}p_{3\nu}-p_{3\mu}p_{1\nu})\nonumber\\&&
  -q^2(\xi y^2-4yz+2\xi yz+2y^2z+\xi z^2+4yz^2)
  (p_{1\mu}s_\nu-s_\mu p_{1\nu})\nonumber\\&&
  -2q^2yz^2(p_{2\mu}s_\nu-s_\mu p_{2\nu})
  -q^2(\xi y^2-2yz+\xi yz+2y^2z)(p_{3\mu}s_\nu-s_\mu p_{3\nu})
  \ \Bigg]
\end{eqnarray}
and\goodbreak
\begin{eqnarray}
H_{\mu\nu}^4&=&\frac{2ig_s^2\NC\CF}{y^2z^2q^4}\Bigg[ 
  -q^2(\xi y^2-4yz+2\xi yz-2y^2z+\xi z^2+2yz^2)\eps(\mu\nu p_1p_2)
  \nonumber\\&&
  -q^2(2y^2-\xi y^2-2yz+2\xi yz+\xi z^2)\eps(\mu\nu p_1p_3)
  +4q^2\xi y^2\eps(\mu\nu p_2p_3)\nonumber\\&&
  +8y(yp_{1\mu}+zp_{2\mu}+yp_{3\mu})\eps(\nu p_1p_2p_3)
  -8y(yp_{1\nu}+zp_{2\nu}+yp_{3\nu})\eps(\mu p_1p_2p_3)\ \Bigg]
  \nonumber\\&& +\frac{2mg_s^2\NC\CF}{y^2z^2q^4}\Bigg[
  -q^2((\xi y^2-4yz+2\xi yz+2y^2z+\xi z^2+2yz^2)(p_2s)
  \nonumber\\&&\qquad\qquad
  +(-2y^2+\xi y^2-2yz+2\xi yz+4y^2z+\xi z^2)(p_3s))g_{\mu\nu}\nonumber\\&&
  -4y^2(p_3s)(p_{1\mu}p_{2\nu}+p_{2\mu}p_{1\nu})
  +8yz(p_3s)p_{2\mu}p_{2\nu}\nonumber\\&&
  -4y(z(p_2s)+y(p_3s))(p_{2\mu}p_{3\nu}+p_{3\mu}p_{2\nu})
  +2q^2y^2z(p_{1\mu}s_\nu+s_\mu p_{1\nu})\nonumber\\&&
  +q^2(\xi y^2-4yz+2\xi yz+4y^2z+\xi z^2+2yz^2)(p_{2\mu}s_\nu+s_\mu p_{2\nu})
  \nonumber\\&&
  +q^2(-2yz+\xi yz+2y^2z+\xi z^2)(p_{3\mu}s_\nu+s_\mu p_{3\nu})
  \ \Bigg],
\end{eqnarray}
where we have accounted for the spin dependence of the hadron tensor by 
using the quark's spin projector
$u(p_1,s)\bar u(p_1,s)=(\slp_1+m)\frac12(1+\gamma_5\sls)$ when 
calculating the trace according to Eq.~(\ref{eqn1}). We have used the 
energy variables $y=1-2p_1\cdot q/q^2$ and $z=1-2p_2\cdot q/q^2$ and the 
abbreviation $\xi=4m^2/q^2$. Note that for the spin dependent 
contributions, $H_{\mu\nu}^{1,2,3}(s)$ are antisymmetric in $\mu$ and 
$\nu$, while $H_{\mu\nu}^4(s)$ is symmetric in $\mu$ and $\nu$. 
For the spin independent pieces, the nonvanishing contributions
$H_{\mu\nu}^{1,2}$ are symmetric, $H_{\mu\nu}^4$ is antisymmetric in~$\mu$ 
and~$\nu$, and there is no spin independent contribution to $H_{\mu\nu}^3$.

In this paper we are only concerned with the longitudinal polarization. 
The covariant form of the polarization four-vector associated with the 
longitudinal polarization is given by
$s^{\ell\mu}=\xi^{-1/2}(\sqrt{(1-y)^2-\xi},0,0,1-y)$ 
($s^\ell_\mu s^{\ell\mu}=-1$) which reduces to $s^{\ell\mu}=(0,0,0,1)$ in 
the rest system of the quark when $y=1-\sqrt\xi$. We then define unpolarized 
and polarized structure functions $H_{\mu\nu}^i$ and $H_{\mu\nu}^{i\ell}$ 
($i=1,2,3,4$) according to
\begin{eqnarray}
H_{\mu\nu}^i&=&H_{\mu\nu}^i(s^\ell)+H_{\mu\nu}^i(-s^\ell),\label{eqn3}\\
H_{\mu\nu}^{i\ell}&=&H_{\mu\nu}^i(s^\ell)-H_{\mu\nu}^i(-s^\ell).\label{eqn4}
\end{eqnarray}
In order to determine the polar angle dependence of the polarized and 
unpolarized cross section one turns to the helicity structure functions 
$H_U^{i(\ell)}=H_{++}^{i(\ell)}+H_{--}^{i(\ell)}$, 
$H_L^{i(\ell)}=H_{00}^{i(\ell)}$ and 
$H_F^{i(\ell)}=H_{++}^{i(\ell)}-H_{--}^{i(\ell)}$ which can be 
obtained from the covariant structure functions in Eqs.~(\ref{eqn3}) 
and~(\ref{eqn4}) by the appropiate helicity projections of the gauge boson,
\begin{equation}\label{eqn5}
H_{\lambda_{Z,\gamma}\lambda_{Z,\gamma}}^{i(\ell)}=\eps^\mu(\lambda_{Z,\gamma})
  H_{\mu\nu}^{i(\ell)}\eps^{\nu*}(\lambda_{Z,\gamma}).
\end{equation}
In Eq.~(\ref{eqn5}) we have introduced a compact bracket notation $(\ell)$ 
which considerably simplifies the presentation in the remaining part of the 
paper. The notation is such that e.g. $H_{\mu\nu}^{i(\ell)}$ stands for 
either the unpolarized hadron tensor $H_{\mu\nu}^i$ or the polarized 
hadron tensor $H_{\mu\nu}^{i\ell}$.

A convenient way of obtaining the helicity structure functions 
$H_\alpha^{i(\ell)}$ ($\alpha=U,L,F$) is by covariant projection. One has
\begin{equation}\label{eqn6}
H_U^{i(\ell)}=\left(-\hat g^{\mu\nu}
  -\frac{\hat p_1^\mu\hat p_1^\nu}{p_{1z}^2}\right)H_{\mu\nu}^{i(\ell)}\qquad
H_L^{i(\ell)}=\frac{\hat p_1^\mu\hat p_1^\nu}{p_{1z}^2}
  H_{\mu\nu}^{i(\ell)}\qquad
H_F^{i(\ell)}=i\eps^{\mu\nu\alpha\beta}
  \frac{\hat p_{1\alpha}q_\beta}{p_{1z}\sqrt{q^2}}H_{\mu\nu}^{i(\ell)},\qquad
\end{equation}
where $\hat g_{\mu\nu}=g_{\mu\nu}-q_\mu q_\nu/q^2$ and
$\hat p_{1\mu}=p_{1\mu}-(p_1\cdot q)q_\mu/q^2$ are the four-transverse 
metric tensor and the four-transverse quark momentum, respectively. Note 
that the covariant projectors defined in Eq.~(\ref{eqn6}) are 
$(y,z)$-independent and can thus be freely commuted with the $y$- and 
$z$-integrations to be done later on.

After all these preliminaries let us write down the differential polarized 
and unpolarized three-body $e^+e^-$-cross sections, differential in $\ct$ 
and in the two energy variables $y$ and $z$. One has
\begin{eqnarray}
\frac{d\sigma^\ell}{d\ct\,dy\,dz}
  &=&\frac38(1+\cos^2\theta)g_{14}\frac{d\sigma_U^{4\ell}}{dy\,dz}
  +\frac34\sin^2\theta\ g_{14}\frac{d\sigma_L^{4\ell}}{dy\,dz}\nonumber\\&&
  +\frac34\cos\theta\left(g_{41}\frac{d\sigma_F^{1\ell}}{dy\,dz}
  +g_{42}\frac{d\sigma_F^{2\ell}}{dy\,dz}\right)\qquad\mbox{and}
  \label{eqn7}\\
\frac{d\sigma}{d\ct\,dy\,dz}
  &=&\frac38(1+\cos^2\theta)\left(g_{11}\frac{d\sigma_U^1}{dy\,dz}
  +g_{12}\frac{d\sigma_U^2}{dy\,dz}\right)\nonumber\\&&
  +\frac34\sin^2\theta\left(g_{11}\frac{d\sigma_L^1}{dy\,dz}
  +g_{12}\frac{d\sigma_L^2}{dy\,dz}\right)
  +\frac34\ct\ g_{44}\frac{d\sigma_F^4}{dy\,dz}\label{eqn8}
\end{eqnarray}
where, in terms of the helicity structure functions defined above, one has 
($\alpha=U,L,F$)
\begin{equation}
\frac{d\sigma_\alpha^{i(\ell)}}{dy\,dz}=\frac{\pi\alpha^2v}{3q^4}\left\{
  \frac{q^2}{16\pi^2v}H_\alpha^{i(\ell)}\right\},
\end{equation}
where $v=\sqrt{1-4m^2/q^2}$ is the velocity of the quark in the 
c.m.~system. The $g_{ij}$ coupling factors appearing in Eqs.~(\ref{eqn7}) 
and~(\ref{eqn8}) specify the electroweak structure of the lepton-hadron 
interaction and are listed in Appendix~C. The $y$, $z$ and $\ct$-dependent 
polarization $\Pl(\ct,y,z)$ is then given by the ratio of the polarized 
and unpolarized cross sections in Eqs.~(\ref{eqn7}) and~(\ref{eqn8}).

The generalization of the above cross section expressions to the case where 
one starts with longitudinally polarized beams is straightforward and 
amounts to the replacement\\
{\it polarized:}
\vspace{-1cm}
\begin{eqnarray}
g_{14}&\rightarrow&[(1-h^-h^+)g_{14}+(h^--h^+)g_{44}]\\
g_{4i}&\rightarrow&[(h^--h^+)g_{1i}+(1-h^-h^+)g_{4i}]\quad(i=1,2)\nonumber
\end{eqnarray}
\noindent{\it unpolarized:}
\vspace{-1cm}
\begin{eqnarray}
g_{1i}&\rightarrow&[(1-h^-h^+)g_{1i}+(h^--h^+)g_{4i}]\quad(i=1,2)\\
g_{44}&\rightarrow&[(h^--h^+)g_{14}+(1-h^-h^+)g_{44}]\nonumber
\end{eqnarray}
where $h^-$ and $h^+$ ($-1\le h^\pm\le+1$) denote the longitudinal 
polarization of the electron and the positron beam. Clearly there is no 
interaction between the beams when $h^+=h^-=\pm1$.

We have presented the results of calculating the full ($U+L$) and the 
longitudinal component ($L$) of the polarized hadron tensor 
in~\cite{forward1,forward3}. The unpolarized transverse component~($U$) can 
then be obtained by subtraction, 
$H_U^{i(\ell)}=H_{U+L}^{i(\ell)}-H_L^{i(\ell)}$. Here we add the last 
building block, namely the polarized hadron tensor projected on its 
forward/backward component~($F$), where the requisite covariant projector 
has been given in Eq.~(\ref{eqn6}). A straightforward calculation of the 
$O(\as)$ tree-graph contributions leads to
\begin{eqnarray}
H_F^{1\ell}(y,z)&=&\frac{8\pi\as\NC\CF}{(1-y)^2-\xi}\Bigg[\
  8-4\xi-\xi^2-2(2-\xi)(2-3\xi)\frac1y\nonumber\\&&
  -\xi(1-\xi)(2-\xi)\left(\frac1{y^2}+\frac1{z^2}\right)
  -2\xi y-4yz-6\xi z-4(2-\xi)(3-2\xi)\frac1z\nonumber\\&&
  +2(1-\xi)(2-\xi)^2\frac1{yz}+(4-2\xi-\xi^2)\frac zy
  +(28-14\xi+\xi^2)\frac yz\nonumber\\&&
  +2\xi(2-\xi)\frac y{z^2}-\xi(4-\xi)\frac{y^2}{z^2}+4\frac{y^3}z
  +2\xi\frac{y^3}{z^2}-2(8-3\xi)\frac{y^2}z\Bigg]
  \label{eqn9}\\[12pt]
H_F^{2\ell}(y,z)&=&\frac{8\pi\as\NC\CF}{(1-y)^2-\xi}\ \xi\Bigg[\
  4+\xi-2(2-3\xi)\frac1y-\xi(1-\xi)\left(\frac1{y^2}+\frac1{z^2}\right)
  \nonumber\\&&
  +4z-4(3-2\xi)\frac1z+2(1-\xi)(2-\xi)\frac1{yz}+\xi\frac zy
  \nonumber\\&&
  +(12-\xi)\frac yz+2\xi\frac y{z^2}-\xi\frac{y^2}{z^2}-4\frac{y^2}z\Bigg].
\end{eqnarray}
For the sake of completeness, we also include results for the unpolarized 
forward/backward ($F$) structure function
\begin{eqnarray}
H_F^4(y,z)&=&\frac{16\pi\as\NC\CF}{\sqrt{(1-y)^2-\xi}}
  \Bigg[-(4-5\xi)\frac1y-\xi(1-\xi)\left(\frac1{y^2}+\frac1{z^2}\right)
  \nonumber\\&&
  -2(4-3\xi)\frac1z+\xi\frac y{z^2}+2z+2\frac zy+6\frac yz-2\frac{y^2}z
  +2(1-\xi)(2-\xi)\frac1{yz}\Bigg].
  \label{eqn10}
\end{eqnarray}
In all three above expressions the denominator vanishes when the quark's 
three-momentum is zero, i.e.\ when $\vec p_1=0$ or, in terms of the 
$y$-variable, when $y=1-\sqrt\xi$. However, a careful limiting procedure 
shows that the three forward/backward structure functions in 
Eqs.~(\ref{eqn9})--(\ref{eqn10}) tend to finite limiting values (and not 
to zero as in the case of the longitudinal structure functions $H_L^{i\ell}$ 
treated in~\cite{forward3}). In contrast to this, the singularities at 
$y=z=0$ constitute true IR-singularities. They can be regularized by 
introducing a small gluon mass~$m_g=\sqrt{\Lambda q^2}$ which has the effect 
to deform the phase-space boundary to
\begin{eqnarray}
y_-&=&\sqrt{\Lambda\xi}+\Lambda,\qquad y_+\ =\ 1-\sqrt\xi\\
z_\pm(y)&=&\frac{2y}{4y+\xi}\left\{1-y-\frac{1}{2}\xi+\Lambda+\frac{\Lambda}{y}
  \pm\frac{1}{y}\sqrt{(y-\Lambda)^2-\Lambda\xi}\sqrt{(1-y)^2-\xi}\right\}.
\end{eqnarray}
Note that the introduction of a small gluon mass is required only to deform 
the integration boundary. In the calculation of the matrix elements we need 
not pay regard to the gluon mass. With the gluon-mass regulator one can then 
perform the finite integration over the two phase-space variables $y$ and $z$. 
The integration over $z$ gives the structure function's dependence on the 
quark energy variable $y$. We obtain 
($H_\alpha^{i(\ell)}(y):=\int dz\,H_\alpha^{i(\ell)}(y,z)$)
\begin{eqnarray}
H_F^{1\ell}(y)&=&\frac{8\pi\as\NC\CF}{(1-y)^2-\xi}\Bigg[\ 2\Bigg\{
  2\frac{(1-\xi)(2-\xi)^2}y-4(3-2\xi)(2-\xi)\nonumber\\&&\qquad\qquad
  +(28-14\xi+\xi^2)y-2(8-3\xi)y^2+4y^3\Bigg\}
  \ln\left(\frac{z_+(y)}{z_-(y)}\right)\nonumber\\&&
  +\frac1y\sqrt{(y-\Lambda)^2-\Lambda\xi}\sqrt{(1-y)^2-\xi}\Bigg\{
  -\frac{8\xi(1-\xi)(2-\xi)y}{\xi y^2+4\Lambda(1-y-\xi+\Lambda)}
  \nonumber\\&&\qquad\qquad
  -8\frac{(1-\xi)(2-\xi)}y+32(2-\xi)-2(18-5\xi)y+20y^2
  \nonumber\\&&\qquad\qquad
  -\frac{(4-\xi)(40+3\xi)y}{2(4y+\xi)}+\frac{(4-\xi)^3y}{(4y+\xi)^2}\Bigg\}
  \Bigg]\\[12pt]
H_F^{2\ell}(y)&=&\frac{8\pi\as\NC\CF}{(1-y)^2-\xi}\ \xi\Bigg[\ 2\Bigg\{
  2\frac{(1-\xi)(2-\xi)}y-4(3-2\xi)\nonumber\\&&\qquad\qquad
  +(12-\xi)y-4y^2\Bigg\}\ln\left(\frac{z_+(y)}{z_-(y)}\right)\nonumber\\&&
  +\frac1y\sqrt{(y-\Lambda)^2-\Lambda\xi}\sqrt{(1-y)^2-\xi}\Bigg\{
  -\frac{8\xi(1-\xi)y}{\xi y^2+4\Lambda(1-y-\xi+\Lambda)}
  \nonumber\\&&\qquad\qquad
  -8\frac{1-\xi}y+32-12y-4\frac{(4-\xi)y}{4y+\xi}\Bigg\}
  \Bigg]\\[12pt]
H_F^4(y)&=&\frac{16\pi\as\NC\CF}{\sqrt{(1-y)^2-\xi}}\Bigg[\ 2\Bigg\{
  \frac{(1-\xi)(2-\xi)}y-(4-3\xi)+3y-y^2\Bigg\}
  \ln\left(\frac{z_+(y)}{z_-(y)}\right)\nonumber\\&&
  +\frac1y\sqrt{(y-\Lambda)^2-\Lambda\xi}\sqrt{(1-y)^2-\xi}\Bigg\{
  -\frac{4\xi(1-\xi)y}{\xi y^2+4\Lambda(1-y-\xi+\Lambda)}
  \nonumber\\&&\qquad\qquad
  -4\frac{1-\xi}y+8-y-\frac{16y}{4y+\xi}+\frac{(4-\xi)^2y}{(4y+\xi)^2}\Bigg\}
  \Bigg].
\end{eqnarray}

\begin{figure}\begin{center}
\epsfig{figure=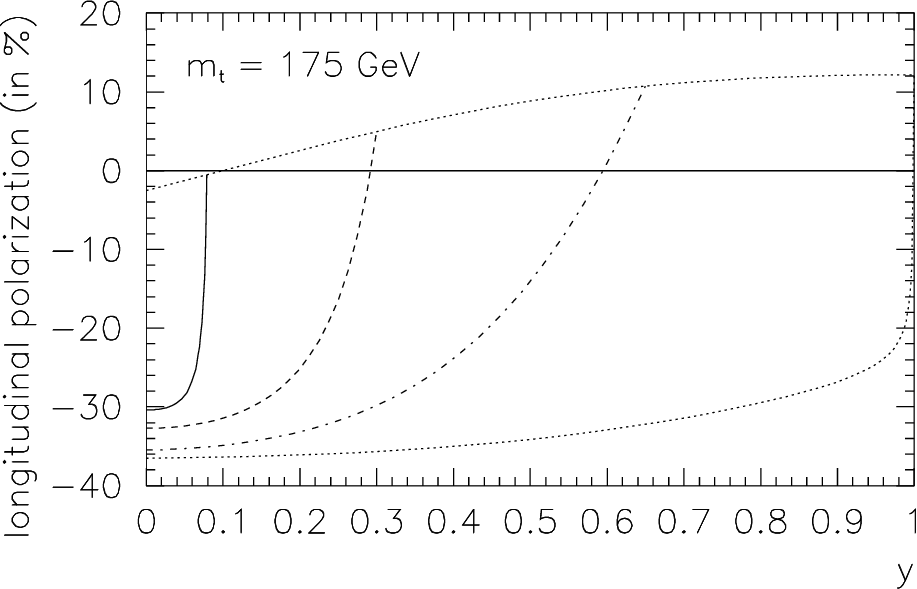, scale=0.8}
\caption{\label{fig2}$y$-dependence of the forward-backward component
$\Pl_F(y)$ of the longitudinal polarization of top quarks produced in the
continuum for  c.m.~energies of $380$~GeV (solid), $500$~GeV (dashed),
$1000$~GeV (dashed-dotted) and $1000$~TeV (dotted). Also shown is the limiting
curve of $\Pl_F(y)$ for maximal values of $y$ (dotted curve in the upper half
of the figure)}
\end{center}\end{figure}

Let us briefly pause to discuss the $y$-dependence of the forward/backward 
($F$) component of the quark's longitudinal polarization $\Pl_F$ which we 
define as
\begin{equation}\label{eqn11}
\Pl_F=\frac{g_{41}H_F^{1\ell}+g_{42}H_F^{1\ell}}%
  {g_{11}H_{U+L}^1+g_{12}H_{U+L}^2}.
\end{equation}
It can be obtained as the $(2\ct)$-moment of the full $\theta$-dependent 
polarization $\Pl(\ct)$, i.e.\ $\Pl_F=\langle2\ct\,\Pl(\ct)\rangle$. The 
IR-limit $y\rightarrow 0$ in Eq.~(\ref{eqn11}) is well defined since the 
potentially singular term proportional to $1/y$ in the numerator is 
cancelled by the $1/y$-terms of the denominator giving a finite polarization 
value for $y\rightarrow 0$. In fact the limiting value of $\Pl_F$ can be 
calculated to be
\begin{equation}\label{eqn12}
\Pl_F(y\rightarrow 0)
  =\frac{2(2-\xi)g_{41}+2\xi g_{42}}{(4-\xi)g_{11}+3\xi g_{12}}.
\end{equation}
Turning to the other corner of phase-space $y\rightarrow 1-\sqrt\xi$, the 
limiting value of the polarization expressions can be obtained by expanding 
numerator and denominator around $y=1-\sqrt\xi$. As mentioned before, the 
$(U+L)$- and $(L)$-components of the polarization vanish in this limit, 
whereas the $(F)$-component of the polarization tends to the finite limiting 
value
\begin{equation}\label{eqn13}
\Pl_F(y\rightarrow 1-\sqrt\xi)=-\frac{2((2-2\sqrt\xi-\xi)g_{41}
  +\xi g_{42})}{3((4-4\sqrt\xi+3\xi)g_{11}-\xi g_{12})}.
\end{equation}
In Fig.~\ref{fig2} we have plotted the $y$-dependence of $\Pl_F$ for the top
quark  case for the three values $\sqrt{q^2}=380$, $500$ and $1000$~GeV using
a top quark mass of $m_t=180$ GeV~\cite{forward8}.\footnote{We have chosen 
the lowest energy $\sqrt{q^2}=380$ GeV to lie well above the nominal 
threshold value of $\sqrt{q^2}=360$ GeV for a perturbative calculation to 
make sense. As is well known, the production dynamics in the threshold 
region requires the consideration of non-perturbative effects. For a 
discussion of top quark polarization effects in the threshold region 
see~\cite{forward9}.} The polarization $\Pl_F$ can be seen to tend to the 
two limiting values given in Eqs.~(\ref{eqn12}) and~(\ref{eqn13}) as 
$y\rightarrow 0$ and $y\rightarrow 1-\sqrt\xi$, respectively.

We finally obtain the fully integrated three-body structure functions 
$H_\alpha^{i(\ell)}$ by $y$-integration,
\begin{equation}
H_\alpha^{i(\ell)}(tree)=\int dy\,H_\alpha^{i(\ell)}(y)
  =\int dy\,dz\,H_\alpha^{i(\ell)}(y,z).
\end{equation}
For $\alpha=F$ one has
\begin{eqnarray}
H_F^{1\ell}({\it tree})&=&8\pi\as\NC\CF\bigg[\
  (8-4\xi-\xi^2)J_1-2(2-\xi)(2-3\xi)J_2\nonumber\\&&
  -\xi(1-\xi)(2-\xi)(J_3+J_9)
  -2\xi J_4-4J_6-6\xi J_7-4(2-\xi)(3-2\xi)J_8\nonumber\\&&
  +2(1-\xi)(2-\xi)^2J_{10}+(4-2\xi-\xi^2)J_{11}
  +(28-14\xi+\xi^2)J_{12}\nonumber\\&&
  +2\xi(2-\xi)J_{13}-\xi(4-\xi)J_{14}+4J_{15}
  +2\xi J_{16}-2(8-3\xi)J_{17}\bigg]\\[12pt]
H_F^{2\ell}({\it tree})&=&8\pi\as\NC\CF\xi\bigg[\
  (4+\xi)J_1-2(2-3\xi)J_2-\xi(1-\xi)(J_3+J_9)\nonumber\\&&
  +4J_7-4(3-2\xi)J_8+2(1-\xi)(2-\xi)J_{10}+\xi J_{11}\nonumber\\&&
  +(12-\xi)J_{12}+2\xi J_{13}-\xi J_{14}-4J_{17}\bigg]\\[12pt]
H_F^4({\it tree})&=&16\pi\as\NC\CF\bigg[
  -(4-5\xi)S_2-\xi(1-\xi)(S_3+S_5)-2(4-3\xi)S_4\nonumber\\&&
  +\xi S_6+2S_8+2S_9+6S_{10}-2S_{11}+2(1-\xi)(2-\xi)S_{12}\bigg].
\end{eqnarray}
The integrals $S_i$ have been computed and listed in~\cite{forward1}. 
Similar techniques allow one to calculate the integrals $J_i$ (for details 
see~\cite{forward1,forward3}). They are listed in Appendix~A. Remember that 
the IR-singularities in $S_3$, $S_5$, $S_{12}$, $J_3$, $J_9$ and $J_{10}$ 
are cancelled by the corresponding IR-singularities of the one-loop 
contributions as a consequence of the Lee-Nauenberg theorem. However, 
before turning to the full $O(\as)$ results in Sec.~4, we first take a brief 
detour in Sec.~3 and consider the soft gluon region of the phase space.

\section{Soft-gluon region}
For some applications it is desirable to split the three-body phase-space 
into a soft and a hard gluon region. The two regions are defined with 
respect to a fixed cut-off value of the gluon energy 
$E_g^{\mbox{\scriptsize cut-off}}=\lambda\sqrt{q^2}$ such that 
$E_g/\sqrt{q^2}\le\lambda$ and $E_g/\sqrt{q^2}>\lambda$ define the soft and 
the hard-gluon regions, respectively. Technically the integration over the 
soft-gluon region is quite simple. If necessary, the hard-gluon contribution 
can then be obtained as the complement of the soft-gluon contribution. Since 
we provide numerically stable expressions for both the total $O(\as)$ 
contribution and the soft-gluon contribution the computation of the 
hard-gluon contribution as the difference of the two constitutes a 
numerically stable procedure without that differences of large numbers are 
encountered.

In as much as the soft-gluon contribution factorizes into a Born term part 
and a universal soft-gluon function we begin our discussion by listing the 
relevant nonvanishing Born term components of the two-body hadron tensor. 
One has
\smallskip\\
{\it Born term:}
\begin{eqnarray}
H_U^{4\ell}({\it Born})&=&4\NC q^2v\nonumber\\
H_L^{4\ell}({\it Born})&=&0\nonumber\\
H_F^{1\ell}({\it Born})&=&2\NC q^2(1+v^2)\qquad
H_F^{2\ell}({\it Born})\ =\ 2\NC q^2(1-v^2)\nonumber\\
H_U^1({\it Born})&=&2\NC q^2(1+v^2)\qquad
H_U^2({\it Born})\ =\ 2\NC q^2(1-v^2)\\
H_L^1({\it Born})&=&\NC q^2(1-v^2)\ =\ H_L^2({\it Born})\nonumber\\
H_F^4({\it Born})&=&4\NC q^2.\nonumber
\end{eqnarray}
For completeness we note that the polarized and unpolarized Born 
term cross sections can be obtained from Eqs.~(\ref{eqn7}) and~(\ref{eqn8}) 
by the following replacements. On the l.h.s.\ of Eqs.~(\ref{eqn7}) 
and~(\ref{eqn8}) one effects the replacements
\begin{equation}\label{eqn14}
\frac{d\sigma^{(\ell)}}{d\ct\,dy\,dz}\rightarrow
  \frac{d\sigma^{(\ell)}}{d\ct}
\end{equation}
and on the r.h.s.~of Eqs.~(\ref{eqn7}) and~(\ref{eqn8}) one rewrites
\begin{equation}\label{eqn15}
\frac{d\sigma_\alpha^{i(\ell)}}{dy\,dz}\rightarrow
  \frac{\pi\alpha^2v}{3q^4}H_\alpha^{i(\ell)}({\it Born}).
\end{equation}

Turning to the soft-gluon region of the $O(\as)$ tree-graph contribution, 
it is well-known that the soft hadronic tensor factorizes into the Born 
term structure times a universal function of the soft-gluon energy. 
Integrating the soft-gluon spectrum and adding the $O(\as)$ soft-gluon 
contribution to the Born term contribution, one obtains
\begin{eqnarray}
H_\alpha^i&=&H_\alpha^i({\it Born})\Bigg[\ 1-\frac{\as\CF}\pi
  \Bigg\{\ln\left(\frac{2\lambda}{\sqrt\Lambda}\right)
  \left(2+\frac{1+v^2}v\ln\left(\frac{1-v}{1+v}\right)\right)\nonumber\\&&
  +\frac1v\ln\left(\frac{1-v}{1+v}\right)
  +\frac{1+v^2}v\left(\Li\left(\frac{2v}{1+v}\right)
  +\frac14\ln^2\left(\frac{1-v}{1+v}\right)\right)\Bigg\}\Bigg]
  \label{eqn16}
\end{eqnarray}
where the scaled maximal gluon energy $\lambda$ denotes the cut-off energy 
which separates the soft-gluon region from the hard-gluon region. The IR 
singularity present in the soft-gluon region has been regularized as before 
by introducing a small gluon mass $m_g=\sqrt{\Lambda q^2}$ which of course 
can be chosen to be arbitrarily small compared to the energy cut-off 
parameter~$\lambda$.

The total $O(\as)$ soft-gluon contribution is then obtained by adding in 
the one-loop contribution. For the latter one has (see 
e.g.~\cite{forward1,forward3})
\medskip\\
{\it one-loop real part:}
\begin{eqnarray}
H_U^{4\ell}({\it loop})&=&4\NC q^2v(\real A+\real C)\nonumber\\
H_L^{4\ell}({\it loop})&=&0\label{eqn17}\\
H_F^{1\ell}({\it loop})&=&4\NC q^2(\real A+v^2\real C)\qquad
H_F^{2\ell}({\it loop})\ =\ 4\NC q^2(\real A-v^2\real C)\nonumber\\
H_U^1({\it loop})&=&4\NC q^2(\real A+v^2\real C)\qquad
H_U^2({\it loop})\ =\ 4\NC q^2(\real A-v^2\real C)\nonumber\\
H_L^1({\it loop})&=&2\NC q^2((1-v^2)\real A+v^2\real B)
  \ =\ H_L^2({\it loop})\nonumber\\
H_F^4({\it loop})&=&4\NC q^2v(\real A+\real C)\nonumber
\end{eqnarray}
\goodbreak\medskip\noindent
{\it one-loop imaginary part:}
\begin{equation}
H_U^{3\ell}({\it loop})=H_F^3({\it loop})=-4\NC q^2v(\imag A-\imag C)\qquad
H_L^{3\ell}({\it loop})=0,
\end{equation}
where the one-loop form factors $A$, $B$ and $C$ are given by
\begin{eqnarray}
\real A&=&-\frac{\as\CF}{4\pi}\Bigg[\left(2+\frac{1+v^2}v
  \ln\left(\frac{1-v}{1+v}\right)\right)
  \ln\left(\frac{\Lambda q^2}{m^2}\right)
  +3v\ln\left(\frac{1-v}{1+v}\right)+4\nonumber\\&&
  +2\frac{1+v^2}v\left(\Li\left(\frac{2v}{1+v}\right)
  +\frac14\ln^2\left(\frac{1-v}{1+v}\right)-\frac{\pi^2}2\right)\Bigg]
  \nonumber\\
\imag A&=&-\frac{\as\CF}{4\pi}\Bigg[\frac{1+v^2}v\left(
  \left(\frac{\Lambda q^2}{m^2}\right)+\ln\left(\frac{1-v^2}{4v^2}\right)
  \right)+3v\Bigg]\nonumber\\
\real B&=&\frac{\as\CF}{4\pi}\frac{1-v^2}v\ln\left(\frac{1-v}{1+v}\right)
\qquad\real C\ =\ \real A-2\real B\\
\imag B&=&\frac{\as\CF}{4\pi}\frac{1-v^2}v\pi\qquad
\imag C\ =\ \imag A-2\imag B.\nonumber
\end{eqnarray}
The $\gamma_5$-even one-loop results for $H_{U,L}^{1,2}$ and $H_{U,L}^{4\ell}$ 
are taken from~\cite{forward1,forward3}. The $\gamma_5$-odd one-loop 
contributions $H_F^{1\ell}$, $H_F^{2\ell}$ and $H_F^4$ have been calculated 
in dimensional regularization with zero gluon mass using a naive 
anticommuting $\gamma_5$ (see e.g.~\cite{forward19}). The result was then 
converted to the gluon mass regularization scheme with the help of the 
substitution 
$1/\eps-\gamma_E+\ln(4\pi\mu^2/m^2)\rightarrow\ln(\Lambda q^2/m^2)$.

Although the imaginary part of the one-loop contribution is only needed in 
the case of transverse normal polarization (see \cite{forward4}), we have 
included it for completeness. The longitudinal polarization treated in this 
paper obtains only contributions from the real part of the one-loop graphs. 
We shall henceforth omit the specification ``real part'' and denote the 
one-loop contributions by $H_\alpha^{i(\ell)}({\it loop})$. It is quite 
apparent that the sum of the real gluon emission contributions in 
Eq.~(\ref{eqn16}) and the real one-loop contributions in Eq.~(\ref{eqn17}) 
are IR finite, i.e.\ the dependence on $\Lambda$ drops out in the sum.

We have numerically evaluated the polarization in the soft-gluon region as 
a function of the gluon energy cut-off. We did not plot the results since 
the $O(\as)$ corrections to the polarization are quite small. In the scale 
of our figures the corrections only become visible for $\lambda<0.1\%$ 
which is too low a cut-off-value for a fixed order perturbation series to 
make sense. In fact, the unpolarized $O(\as)$ cross section crosses zero and 
becomes negative at around $\lambda=0.05\%$ which again highlights the fact 
that such low cut-off values do not make any sense. One concludes that the 
$O(\as)$ corrections go in the same direction for both polarized and 
unpolarized cross sections rendering the ratio unsensitive to $O(\as)$ 
corrections.

\section{Total \boldmath{$O(\as)$} results}
Let us now present our total polarized $O(\as)$ results. They will be 
written in terms of the $O(\as)$ two-body structure functions 
$H_\alpha^{i(\ell)}(\as)$ which are determined from the sum of the fully 
integrated three-body structure functions $H_\alpha^{i(\ell)}({\it tree})$
and the two-body one-loop structure functions 
$H_\alpha^{i(\ell)}({\it loop})$. The former have to be multiplied by the 
relative three-body/two-body phase-space factor $q^2/16\pi^2v$ when doing 
the sum. One has
\begin{equation}\label{eqn17a}
H_\alpha^{i(\ell)}(\as)
  =\frac{q^2}{16\pi^2v}H_\alpha^{i(\ell)}({\it tree})
  +H_\alpha^{i(\ell)}({\it loop}).
\end{equation}
Note that the dependence on the IR-cutoff parameter $\Lambda$ drops out in 
the sum of the two contributions. For the sake of completeness we list all 
non-vanishing polarized and unpolarized components $H_\alpha^{i\ell}(\as)$ 
and $H_\alpha^i(\as)$ ($\alpha=U,L,F$). One has
\begin{eqnarray}
H_U^{4\ell}(\as)&=&\frac{\as\NC\CF q^2}{4\pi v}\Big[
  -2(1-\sqrt\xi)(2-6\sqrt\xi+29\xi)-16v^2t_{11}-16v^2t_{10}\nonumber\\&&
  +8(2-\xi)v(t_7-t_8)-2(8+2\xi+3\xi^2)t_6+4(4+9\xi)vt_3\nonumber\\&&
  -(16-30\xi+\frac{17}2\xi^2)(t_1-t_2)\ \Big]\label{eqn18}
  \\[12pt]
H_L^{4\ell}(\as)&=&\frac{\as\NC\CF q^2}{4\pi v}\Big[\
  4(1-\sqrt\xi)(2-6\sqrt\xi+13\xi)+2\xi(10+3\xi)t_6\nonumber\\&&
  -52\xi vt_3-\xi(24-7\xi)(t_1-t_2)\ \Big]
  \\[12pt]
H_F^{1\ell}(\as)&=&\frac{\as\NC\CF q^2}{4\pi v}\Big[
  -4(2+3\xi)v-16(2-\xi)vt_{12}-8(2-\xi)vt_{10}\nonumber\\&&
  -4(2-\xi)^2(t_8-t_9)+2\xi(10-\xi)t_5
  +2\sqrt\xi(1-\sqrt\xi)(2-\sqrt\xi)(4+\sqrt\xi)t_4\nonumber\\&&
  +2(24-12\xi+\xi^2)t_3\ \Big]\label{eqn19}
  \\[7pt]
H_F^{2\ell}(\as)&=&\frac{\as\NC\CF q^2}{4\pi v}\xi\Big[\
  12v-16vt_{12}-8vt_{10}-4(2-\xi)(t_8-t_9)\nonumber\\&&
  +2\xi t_5+2\sqrt\xi(1-\sqrt\xi)t_4+2(6-\xi)t_3\ \Big]
  \\[7pt]
H_U^1(\as)&=&\frac{\as\NC\CF q^2}{4\pi v}\Big[\
  2(2+7\xi)v-16(2-\xi)vt_{12}-8(2-\xi)vt_{10}-4(2-\xi)^2(t_8-t_9)\nonumber\\&&
  -2\xi(2+3\xi)t_5+2\sqrt\xi(1-\sqrt\xi)(2+4\sqrt\xi-3\xi)t_4
  +(48-48\xi+7\xi^2)t_3\ \Big]
  \\[7pt]
H_U^2(\as)&=&\frac{\as\NC\CF q^2}{4\pi v}\xi\Big[\
  12v-16vt_{12}-8vt_{10}-4(2-\xi)(t_8-t_9)\nonumber\\&&
  +2\xi t_5+2\sqrt\xi(1-\sqrt\xi)t_4
  +2(6-\xi)t_3\ \Big]
  \\[7pt]
H_L^1(\as)&=&\frac{\as\NC\CF q^2}{4\pi v}\Big[\ (8-23\xi+\frac32\xi^2)v
  -8\xi vt_{12}-4\xi vt_{10}-2\xi(2-\xi)(t_8-t_9)\nonumber\\&&
  +2\xi(2+3\xi)t_5-2\sqrt\xi(1-\sqrt\xi)(2+4\sqrt\xi-3\xi)t_4\nonumber\\&&
  +\xi(22-8\xi+\frac34\xi^2)t_3\ \Big]
  \\[7pt]
H_L^2(\as)&=&\frac{\as\NC\CF q^2}{4\pi v}\xi\Big[\
  \frac32(10-\xi)v-8vt_{12}-4vt_{10}-2(2-\xi)(t_8-t_9)\nonumber\\&&
  -2\xi t_5-2\sqrt\xi(1-\sqrt\xi)t_4
  +(6-4\xi-\frac34\xi^2)t_3\ \Big]
  \\[7pt]
H_F^4(\as)&=&\frac{\as\NC\CF q^2}{4\pi v}\Big[
  -16\sqrt\xi(1-\sqrt\xi)-16v^2t_{11}-16v^2t_{10}\nonumber\\&&
  +8(2-\xi)v(t_7-t_8)-4(4-5\xi)t_6+8(2-3\xi)vt_3-16(t_1-t_2)\ \Big]
\label{eqn20}
\end{eqnarray}
Closed form expressions for the $O(\as)$ rate functions $t_i$ 
($i=1,\ldots,12$) appearing in Eqs.~(\ref{eqn18}--\ref{eqn20}) can be 
found in Appendix~B. They contain the set of basic $I$, $J$, $S$ and $T$ 
integrals calculated in \cite{forward1} ($I,S$), \cite{forward3} ($T$) and 
in Appendix~A ($J$). They are associated with the various hadron tensor 
components in the following way: $U(S,T)$, $L(T)$, $[U+L](S)$ and $F(J)$ 
(polarized case) and $U(I,J)$, $L(J)$, $[U+L](I)$ and $F(S)$ (unpolarized 
case). We mention that, concerning the unpolarized $(F)$-component in 
Eq.~(\ref{eqn20}), closed form expressions for this component are also 
given in \cite{forward10} and \cite{forward11}. We agree with the results 
of these authors.

The $O(\as)$ structure functions $H_\alpha^{i(\ell)}(\as)$ written down in
Eqs.~(\ref{eqn18}--\ref{eqn20}) constitute two-body structure functions. 
The corresponding $\ct$-dependent cross section expressions follow from
Eqs.~(\ref{eqn7}) and~(\ref{eqn8}) with the same replacements as detailed 
in Eqs.~(\ref{eqn14}) and~(\ref{eqn15}) only that one now has 
$H_\alpha^{i(\ell)}(\as)$ instead of $H_\alpha^{i(\ell)}({\it Born})$.

\begin{figure}\begin{center}
\epsfig{figure=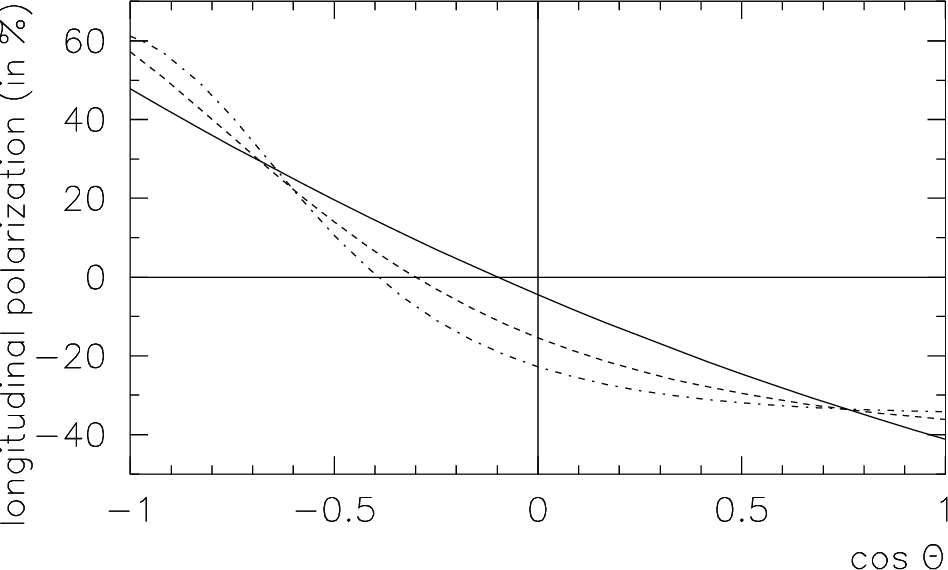, scale=0.8}\\
(a)\\[12pt]
\epsfig{figure=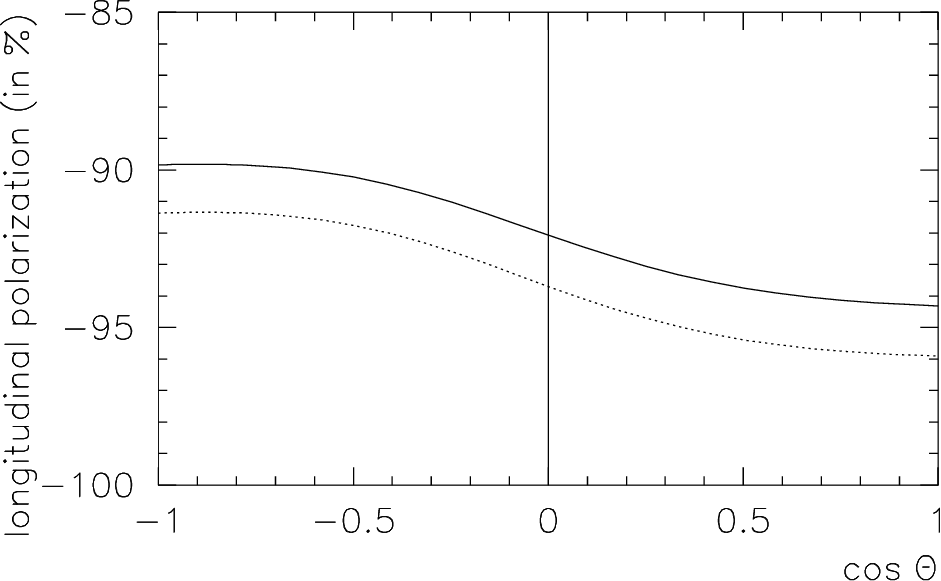, scale=0.8}\\
(b)\\[12pt]
\caption{\label{fig3}Polar angle dependence of the longitudinal polarization 
of (a) top quarks in the continuum for c.m.~energies of $380$~GeV (solid),
$500$~GeV (dashed) and $1000$~GeV (dashed-dotted); (b) bottom quarks on the
$Z$-peak. Shown are Born term results (dotted) and full $O(\as)$ results
(solid)}
\end{center}\end{figure}

For our numerical evaluation we take the current world mean value for $\as$, 
i.e.\ $\as(M_Z)=0.118$ for five active flavours and evolve $\as$ by use of 
the one-loop renormalization group equation. For the bottom and top mass we 
use pole mass values $m_b=4.83$~GeV~\cite{forward12} and 
$m_t=180$~GeV~\cite{forward8}. We have repeated the numerical evaluation 
using $\overline{\mbox{MS}}$ running masses $\bar m_b(\bar m_b)=4.44$~GeV and 
$\bar m_t(\bar m_t)=172.1$~GeV where the running of the masses follows the 
momentum dependence of the corresponding one-loop renormalization group 
equation~\cite{forward13}. However, since the polarization expressions are 
rather insensitive to which sets of masses is used, our numerical results 
in Figs.~\ref{fig2} and~\ref{fig3} are given only in terms of the above pole
masses.

In Fig.~\ref{fig3} we show the polar angle dependence of the longitudinal
polarization of top quarks in the continuum, and for bottom quarks on the $Z$.
We want to emphasize that we define the $O(\as)$ polarization such that we
keep the full $O(\as)$ dependence of the unpolarized cross section in the 
denominator. Thus we do not expand the inverse of the cross section in 
powers of $\as$ as is sometimes done in the literature.

The $\ct$-dependence of the longitudinal polarization of the top quark shown 
in Fig.~\ref{fig3}a is quite strong for all three c.m.~energies $380$~GeV,
$500$~GeV and $1000$~GeV and shows a strong forward-backward asymmetry
signalling a strong $(F)$-component in the polarization. The energy dependence
of the longitudinal polarization is not very pronounced where the asymmetry
becomes somewhat larger as the energy increases. Shown are the full $O(\as)$ 
results for the polarization. We did not plot the corresponding Born term 
results since the respective $O(\as)$ and Born term results differ by less 
than 2\% in absolute value. Again the $O(\as)$ corrections in the numerator 
and in the denominator of the polarization expression tend to go in the 
same direction leaving the polarization practically unchanged under 
$O(\as)$ corrections. For the bottom quark case shown in Fig.~\ref{fig3}b we
plot the Born term results and the full $O(\as)$ results separately. Note, 
however, that the $O(\as)$ corrections amount to only $\cong2\%$ over the 
whole $\ct$ range. In order to be able to exhibit the size of the $O(\as)$ 
corrections we chose to use a ``suppressed zero'' plot in Fig.~\ref{fig3}b.
The $\ct$-dependence of the longitudinal polarization is very weak with a
small asymmetry component. We have checked that using a running mass 
$\bar m_b(\MZ)=3.30$~GeV in Fig.~\ref{fig3}b changes the results by less than
0.2\%.

\section{Massless QCD and the zero-mass limit of QCD}
It is well known by now that, as concerns spin-flip contributions induced 
by gluon radiation, the zero-mass limit of QCD and massless QCD do not 
coincide, at least in the perturbative 
sector~\cite{forward1,forward2,forward3,forward14,forward15}. Since 
familiarity with this statement is not so widespread we want to use this 
section to highlight the difference between the $m=0$ and 
$m\rightarrow 0$ $O(\as)$ results for the longitudinal polarization. 
Needless to say that this has important practical implications for the 
calculation of polarization observables for the process 
$Z\rightarrow\mbox{\it light quark-pairs}$ where one might be tempted to set 
the quark mass to zero from the very beginning because of the presence of 
the large $Z$-mass scale.

Let us begin our discussion by listing the zero-mass Born term 
contributions in terms of the two-body helicity structure functions. 
These are
\begin{eqnarray}
H_U^{4\ell}&=&H_U^1\ =\ H_F^{1\ell}\ =\ H_F^4\ =\ 4\NC q^2\\
H_L^{4\ell}&=&H_L^1\ =\ 0.
\end{eqnarray}

Next we write down the $O(\as)$ zero-mass tree-graph contributions, which 
read
\begin{eqnarray}
H_{U+L}^{4\ell}(y,z)&=&H_{U+L}^1(y,z)
  \ =\ 32\pi\as\NC\CF\ \frac{(1-y)^2+(1-z)^2}{yz}\label{eqn21}\\
H_L^{4\ell}(y,z)&=&H_L^1(y,z)
  \ =\ 32\pi\as\NC\CF\ \frac{2(1-y-z)}{(1-y)^2}\\
H_F^{1\ell}(y,z)&=&H_F^4(y,z)
  \ =\ 32\pi\as\NC\CF\left(\frac{(1-y)^2+(1-z)^2}{yz}-2\,\frac{1-z}{1-y}
  \right).\label{eqn22}
\end{eqnarray}
Because of the $\gamma_5$-invariance property of massless QCD one has 
$H_\alpha^2=H_\alpha^{2\ell}=0$ and $H_\alpha^3=H_\alpha^{3\ell}=0$ to any 
order in~$\as$. This is implicit in the following discussion.

The hadron tensors $H_\alpha$ ($\alpha=U+L,F$) become singular for 
$y\rightarrow 1$ and $z\rightarrow 1$. These singularities need to be 
regularized when integrating over $(y,z)$ phase space. A particularly 
convenient regularization scheme is regularization by dimensional 
reduction~\cite{forward16} where the spin degrees of freedom are kept in 
four dimensions and only scalar integrals are regularized dimensionally. 
In this scheme one can retain the matrix element expressions 
Eq.~(\ref{eqn21}--\ref{eqn22}) when doing the $n$-dimensional phase space 
integration. After adding in the loop contribution (also regularized by 
dimensional reduction) the finite result can be transposed to the 
dimensional regularization scheme by adding in a finite global 
counter-term~\cite{forward17} which will be specified later on. The 
dimensional reduction calculation of the cross sections proceeds along the 
lines discussed in~\cite{forward18}.

The tree-graph expressions have to be integrated with respect to the 
$n$-dimensional integration measure ($n=4-2\eps$)
\begin{equation}
\left(\frac{4\pi\mu^2}{q^2}\right)^\eps
  \frac1{\Gamma(1-\eps)}\int_0^1dy\int_0^{1-y}dz\,y^{-\eps}z^{-\eps}
  (y+z-1)^{-\eps}.
\end{equation}

The integrations can be done in standard fashion, and one obtains the 
tree-graph contributions
\begin{eqnarray}
H_{U+L}^{4\ell}({\it tree})&=&H_{U+L}^1({\it tree})
  \ =\ 64\pi\as\NC\CF C\left(\frac1{\eps^2}
  +\frac3{2\eps}+\frac{17}4\right)\\
H_L^{4\ell}({\it tree})&=&H_L^1({\it tree})
  \ =\ 64\pi\as\NC\CF C\left(\frac12\right)\\
H_F^{1\ell}({\it tree})&=&H_F^4({\it tree})
  \ =\ 64\pi\as\NC\CF C\left(\frac1{\eps^2}
  +\frac3{2\eps}+\frac{17}4-\frac34\right)\label{eqn23}
\end{eqnarray}
where
\begin{equation}
C=\left(\frac{4\pi\mu^2}{q^2}\right)^\eps
  \frac{\Gamma^2(1-\eps)}{\Gamma(1-3\eps)}.
\end{equation}
The right-most finite contributions in the round bracket of 
Eq.~(\ref{eqn23}) results from the integration of the right-most term in 
the round bracket of Eq.~(\ref{eqn22}).

The loop contributions in dimensional reduction are given by
($H_L^{4\ell}=H_L^1=0$)~\cite{forward18}
\begin{equation}\label{eqn24}
H_U^{4\ell}=H_U^1=H_F^{1\ell}=H_F^4=4\NC q^2\frac{\as}\pi\CF C'\left(
  -\frac1{\eps^2}-\frac3{2\eps}-3\right)
\end{equation}
where one has to remember to do all $\gamma$-matrix manipulations in four 
dimensions. $C'$ is given by
\begin{equation}
C'=\left(\frac{4\pi\mu^2}{-q^2}\right)^\eps
  \frac{\Gamma(1+\eps)\Gamma^2(1-\eps)}{\Gamma(1-2\eps)}.
\end{equation}
For the real parts that are of interest here we have $C'=C+O(\eps^3)$. When 
combining the tree-graph contributions with the loop contributions one has 
to remember to include the relative two-body/three-body phase-space factor 
in the former as in Eq.~(\ref{eqn17a}). For massless quarks ($v=1$) the 
relative phase-space factor is given by $q^2/16\pi^2$.

In order to convert our results to the more familiar dimensional 
regularization result we have to add the global counter-term, which reads
($H_L^{4\ell}=H_L^1=0$) \cite{forward17}
\begin{equation}
H_U^{4\ell}=H_U^1=H_F^{1\ell}=H_F^4=4\NC q^2\frac{\as}\pi\CF
  \left(-\frac12\right).
\end{equation}

By adding up all the contributions including the counter-term we arrive at
the full $O(\as)$ result
\begin{eqnarray}
H_U^{4\ell}(\as)&=&H_U^1(\as)
  \ =\ 4\NC q^2\frac{\as}\pi\CF\left(\frac14\right)\label{eqn25}\\
H_L^{4\ell}(\as)&=&H_L^1(\as)
  \ =\ 4\NC q^2\frac{\as}\pi\CF\left(\frac12\right)\\
H_F^{1\ell}(\as)&=&H_F^4(\as)\ =\ 0.\label{eqn26}
\end{eqnarray}
For easy reference one may add the Born term contribution to obtain the 
familiar result ($\CF=4/3$) $H_{U+L}^1=4\NC q^2(1+\as/\pi)$. The vanishing 
of $H_F^4$ and thereby $H_F^{1\ell}$ in massless QCD has been noted 
before~\cite{forward7,forward19,forward20}.\footnote{For a careful 
discussion of the $\gamma_5$-subtleties in dimensional regularization when 
deriving the zero result $H_F^4(\as)=0$ we refer the reader 
to~\cite{forward19,forward20}. In dimensional reduction the troublesome 
$\gamma_5$-odd traces can be completely avoided by deriving the cross section 
expression~Eq.(\ref{eqn22}) and the one-loop contribution~Eq.(\ref{eqn24}) 
in a well-defined manner by use of e.g. helicity amplitudes in which the 
massless axial vector current helicity amplitudes are related to the massless 
vector current helicity amplitudes in canonical fashion. This is implicit 
in Eq.~(\ref{eqn24}) where we have used the relation 
$H_F^{1\ell}=H_F^4=H_U^1=H_U^{4\ell}$ to relate the troublesome 
$\gamma_5$-odd objects $H_F^{1\ell}=H_F^4$ to the unproblematic 
$\gamma_5$-even objects $H_U^{4\ell}$, $H_U^1$ using parity and 
$\gamma_5$-invariance.}

As it turns out one does not reproduce the polarized result in 
Eq.~(\ref{eqn25}) and in Eq.~(\ref{eqn26}) when taking the $m\rightarrow 0$ 
limit in the corresponding expressions Eqs.~(\ref{eqn18}) and~(\ref{eqn19}) 
in Sec.~3, respectively. One obtains instead
\begin{eqnarray}
H_U^{4\ell}(\as)&=&4\NC q^2\frac{\as}\pi\CF
  \left(\frac14-\left[\frac12\right]\right)\label{eqn27}\\
H_F^{4\ell}(\as)&=&4\NC q^2\frac{\as}\pi\CF\left(\frac12\right)\\
H_F^{1\ell}(\as)&=&4\NC q^2\frac{\as}\pi\CF
  \left(0-\left[\frac12\right]\right)\label{eqn28}
\end{eqnarray}
where, by hindsight, we have split the contributions in the round brackets 
into a normal spin non-flip piece and an anomalous spin-flip piece. Note 
the drastic changes in the $O(\as)$ results for $H_U^{4\ell}$ and 
$H_F^{1\ell}$ when the anomalous pieces are included.

The anomalous spin-flip contributions have their origin in the collinear 
limit where the spin-flip contribution proportional to $m$ survives since it 
is multiplied by the $1/m$ collinear mass singularity. Because the 
anomalous spin-flip terms are associated with the collinear singularity, the 
flip contributions are universal and factorize into the Born term 
contribution and an universal spin-flip bremsstrahlung 
function~\cite{forward15}. This explains why there is no anomalous 
contribution to $H_L^{4\ell}$ and why the anomalous flip contributions to 
$H_U^{4\ell}$ and $H_F^{1\ell}$ are equal. In fact the strength of the 
anomalous spin-flip contribution can directly be calculated from the 
universal helicity-flip bremsstrahlung function listed in~\cite{forward15}. 
One has
\begin{equation}
\frac{dH_{hf}}{dy}=H({\it Born})\frac{\as}{2\pi}\CF y
\end{equation}
where $y$ is the energy fraction of the quark after the radiative flip 
transition relative to that of the radiating quark ($0\le y\le 1$). 
Integrating over $y$ one finds
\begin{equation}
\int_0^1dy\ \frac{dH_{hf}}{dy}=H({\it Born})\frac{\as}{2\pi}\CF\,\frac12.
\end{equation}
Then considering the fact that the flip contribution contributes negatively 
to the longitudinal polarization (see Eq.~(\ref{eqn4})), and that there can be
a flip on the antiquark side also, one reproduces exactly the anomalous 
flip contributions in Eqs.~(\ref{eqn27}) and~(\ref{eqn28}).

It is important to realize that the anomalous spin flip contribution does 
not come into play when calculating the transverse polarization of a single 
quark, because of the extra $m/\sqrt{q^2}$ factor in the transverse 
polarization expressions. However, when calculating quark-antiquark 
polarization correlation effects, the $O(\as)$ anomalous contribution does 
come in even when considering transverse polarization effects.

\section{Summary and Conclusion}
We have presented the results of an $O(\as)$ calculation of the full polar 
angle dependence of the quark's longitudinal polarization with respect to the 
electron beam direction including beam polarization effects. This completes 
a series of papers devoted to the full calculation of quark polarization 
effects on and off the $Z$-peak in $e^+e^-$-annihilations into a pair of 
massive quarks. When averaged over phase-space, the $O(\as)$ corrections to 
the polarization observables are generally small (typically of the order 
$2\%$) indicating that the $O(\as)$ corrections to the polarized and 
unpolarized cross sections tend to go in the same direction. This of course 
need not be true everywhere in phase-space and for every polarization 
component. We mention that our results are also of relevance for the 
final-state radiative QED corrections to lepton polarization in 
$e^+e^-\rightarrow\tau^+\tau^-$~\cite{forward21}, where, to our knowledge, 
a full $O(\alpha)$ calculation including quark mass effects has not yet 
been published.

We have not discussed how to measure the polarization of the quarks either 
when it decays as a free quark as in top quarks, or when it decays as a 
bound quark in e.g. polarized $\Lambda_b$-baryon decays. There exist a long 
list of papers devoted to this subject, among which are the papers listed 
in~\cite{forward22,forward23}. Of particular interest is the transverse 
normal polarization studied in~\cite{forward4,forward23,forward24,forward25}. 
It provides the SM background to possible $CP$-violating contributions to 
transverse polarization considered in~\cite{forward24,forward25}. In fact, 
by comparing the transverse normal polarization of the quark and the 
antiquark, the truly $CP$-violating contributions can be 
separated~\cite{forward25}.

Recently there has been a thorough numerical study of gluonic corrections 
to top quark pair production {\em and\/} decay (interference effects not 
included)~\cite{forward26}. By using helicity amplitudes for both the 
production and decay process, the full spin momentum correlation structure 
of the production and decay process has been retained in the calculation 
of~\cite{forward26}. The production-side information in~\cite{forward26} 
is identical to the information contained in our paper. However, we believe 
that our results are physically more transparent since we have chosen a 
two-step approach in which we first represent the production-side dynamics 
in terms of the polarization components of the produced top quarks 
including beam polarization effects as in \cite{forward26}. The 
polarization of the produced top quarks can then be probed by its 
subsequent decay. Also, by using analytical methods, we were able to effect 
large cross section cancellations in sensitive areas of phase-space. These 
could lead to numerically unstable results in the purely numerical approach 
of~\cite{forward26} when not properly dealt with. At any rate, by comparing 
the production-side results of~\cite{forward26} and our results one has a 
powerful check on the technically rather involved calculation of gluonic 
corrections to the polarization of quarks in $e^+e^-$-annihilation.

\vspace{1cm}

\noindent{\bf Acknowledgement:} We would like to thank D.H.~Schiller for 
allowing us to freely quote from the unpublished material 
in~\cite{forward27} when writing Appendix~C.

\newpage

\section*{Appendix A}
\setcounter{equation}{0}
\def\theequation{A\arabic{equation}}
The $J$-type integrals appearing in Section~2 can be calculated using the 
substitution $1-y=\sqrt\xi(1+\eta^2)/(1-\eta^2)$ which leads to the 
replacement of the lower boundary $y=0$ by 
$w:=\eta(y=0)=\sqrt{(1-\sqrt\xi)/(1+\sqrt\xi)}$. Remember also that we 
have $v=\sqrt{1-\xi}$.\\
\vspace{12pt}
\begin{eqnarray}
J_1&=&\jint\nonumber\\ 
  &=&2\,{1-\xi\over 4-\xi}\ln\left({1+v\over 1-v}\right)\\[12pt]
J_2&=&\jint{1\over y}\nonumber\\ 
  &=&{6\over 4-\xi}\ln\left({1+v\over 1-v}\right)\\[12pt]
J_3&=&\jint{1\over y^2}\nonumber\\
  &=&{4\over\xi\,v}\Biggl[\,-\ln\Lambda^{1\over 2}
     -\ln\xi+2\ln v+2\ln2-1\,\Biggr]
     -{24\over\xi(4-\xi)}\ln\left({1+v\over 1-v}\right)\\[12pt]
J_4&=&\jint y\nonumber\\
  &=&-\left(1+\Frac{1}{2}\xi-{6\over 4-\xi}\right)
     \ln\left({1+v\over 1-v}\right)-v\\[12pt]
J_5&=&\jint y^2\nonumber\\ 
  &=&\left(2+\Frac{1}{2}\xi+\Frac{1}{8}\xi^2-{6\over 4-\xi}\right)
     \ln\left({1+v\over 1-v}\right)-\Frac{1}{4}(6-\xi)v\\[12pt]
J_6&=&\jint yz\nonumber\\ 
  &=&\Frac{1}{2}\xi\left(-2+\Frac{1}{8}\xi+{9\over 4-\xi}
     -{12\over(4-\xi)^2}\right)\ln\left({1+v\over 1-v}\right)
     +\left(\Frac{3}{4}+\Frac{1}{8}\xi-{2\over 4-\xi}\right)v\\[12pt]
J_7&=&\jint z\nonumber\\ 
  &=&-{3\xi\over 4-\xi}\left(1-{2\over 4-\xi}\right)
     \ln\left({1+v\over 1-v}\right)+{2v\over 4-\xi}\\[12pt]
J_8&=&\jint{\:1\over z}\nonumber\\
  &=&\frac1{\sxi}\Bigg[\,\Li(w)-\Li(-w)
     +\Li\left({2+\sxi\over 2-\sxi}\;w\right)
     -\Li\left(-{2+\sxi\over 2-\sxi}\;w\right)\,\Biggr]\\[12pt]
J_9&=&\jint{1\over z^2}\nonumber\\ 
  &=&{4\over\xi\,v}\Biggl[\,-\ln\Lambda^{1\over 2}-\ln\xi+2\ln v+2\ln2
     -\frac{2-\xi}{2v}\ln\left(\frac{1+v}{1-v}\right)\,\Biggr]\\[12pt]
J_{10}&=&\jint{1\over yz}\nonumber\\
  &=&{1\over 2(1-\xi)}\ln\left(\frac{1+v}{1-v}\right)
     \Biggl[-4\ln\Lambda^{\frac12}-\Frac{5}{2}\ln\xi+5\ln(1+\sxi)
     +4\ln(1-\sxi)\nonumber\\&&\hphantom{-\ln(2+\sxi)+6\ln2\Biggr]}
     -\ln(2+\sxi)+6\ln2\Biggr]\nonumber\\&&
     +{2\over 1-\xi}\Biggl[\,\Li\left(\frac{1+v}2\right)
     -\Li\left(\frac{1-v}2\right)\,\Biggr]
     +{3\over 1-\xi}\Biggl[\,\Li\left(-{2v\over 1-v}\right)
     -\Li\left(\frac{2v}{1+v}\right)\,\Biggr]\nonumber\\&&
     +{1\over\sxi(1-\sxi)}\Bigg[\,\Li(w)-\Li(-w)
     +\Li\left({2+\sxi\over 2-\sxi}\;w\right)
     -\Li\left(-{2+\sxi\over 2-\sxi}\;w\right)\,\Biggr]\nonumber\\&&
     -{1\over 1-\xi}\Bigg[\Li\left({1+w\over 2}\right)
     -\Li\left({1-w\over 2}\right)
     +\Li\left((2+\sxi){1+w\over 4}\right)\nonumber\\&&
     -\Li\left((2+\sxi){1-w\over 4}\right)
     +\Li\left({2\sxi\over(2+\sxi)(1+w)}\right)
     -\Li\left({2\sxi\over(2+\sxi)(1-w)}\right)\Bigg]\nonumber\\ \\
J_{11}&=&\jint{z\over y}\nonumber\\
  &=&\left(-1+{12\over 4-\xi}-{24\over(4-\xi)^2}\right)
     \ln\left({1+v\over 1-v}\right)-\frac{2v}{4-\xi}\\[12pt]
J_{12}&=&\jint{y\over z}\nonumber\\
  &=&\Frac{1}{2}\ln\left(\frac{1+v}{1-v}\right)\Biggl[\,
     \Frac{1}{2}\ln\xi+\ln(2+\sxi)-\ln(1+\sxi)-2\ln2\,\Biggr]\nonumber\\&&
     +{1-\sxi\over\sxi}\Bigg[\,\Li(w)-\Li(-w)
     +\Li\left({2+\sxi\over 2-\sxi}\;w\right)
     -\Li\left(-{2+\sxi\over 2-\sxi}\;w\right)\,\Biggr]\nonumber\\&&
     +\Li\left({1+w\over 2}\right)-\Li\left({1-w\over 2}\right) 
     +\Li\left((2+\sxi){1+w\over 4}\right)
     -\Li\left((2+\sxi){1-w\over 4}\right)\nonumber\\&&
     +\Li\left({2\sxi\over(2+\sxi)(1+w)}\right)
     -\Li\left({2\sxi\over(2+\sxi)(1-w)}\right)\\[12pt]
J_{13}&=&\jint{y\over z^2}\nonumber\\
  &=&\frac2\xi\ln\left({1+v\over 1-v}\right)\\[12pt]
J_{14}&=&\jint{y^2\over z^2}\nonumber\\
  &=&\frac2\xi\left(\,\ln\left({1+v\over 1-v}\right)-2v\,\right)\\[12pt]
J_{15}&=&\jint{y^3\over z}\nonumber\\
  &=&\Frac{1}{2}(3+\xi)\ln\left({1+v\over 1-v}\right)\Biggl[\,
     \Frac{1}{2}\ln\xi+\ln(2+\sxi)-\ln(1+\sxi)-2\ln2\,\Biggr]\nonumber\\&&
     +{(1-\sxi)^3\over\sxi}\Bigg[\,\Li(w)-\Li(-w)
     +\Li\left(\frac{2+\sxi}{2-\sxi}\;w\right)
     -\Li\left(-\frac{2+\sxi}{2-\sxi}\;w\right)\,\Biggr]\nonumber\\&&
     +(3+\xi)\Bigg[\,\Li\left({1+w\over 2}\right)
     -\Li\left({1-w\over 2}\right) 
     +\Li\left((2+\sxi){1+w\over 4}\right)\nonumber\\&&
     -\Li\left((2+\sxi){1-w\over 4}\right)
     +\Li\left(\frac{2\sxi}{(2+\sxi)(1+w)}\right)
     -\Li\left(\frac{2\sxi}{(2+\sxi)(1-w)}\right)\,\Bigg]\nonumber\\&&
     +{40-16\xi+\xi^2\over16}\ln\left(\frac{1+v}{1-v}\right)
     -{26-\xi\over 8}\,v\\[12pt]
J_{16}&=&\jint{y^3\over z^2}\nonumber\\
  &=&\frac{2+\xi}\xi\ln\left({1+v\over 1-v}\right)-\frac{6}{\xi}v\\[12pt]
J_{17}&=&\jint{y^2\over z}\nonumber\\
  &=&\Frac{1}{2}\ln\left({1+v\over 1-v}\right)
     \Biggl[\,\Frac{1}{2}\ln\xi+\ln(2+\sxi)
     -\ln(1+\sxi)-2\ln2\,\Biggr]\nonumber\\&&
     +{(1-\sxi)^2\over\sxi}\Bigg[\,\Li(w)-\Li(-w)
     +\Li\left(\frac{2+\sxi}{2-\sxi}\;w\right)
     -\Li\left(-\frac{2+\sxi}{2-\sxi}\;w\right)\,\Biggr]\nonumber\\&&
     +\Li\left({1+w\over 2}\right)-\Li\left({1-w\over 2}\right) 
     +\Li\left((2+\sxi){1+w\over 4}\right)\nonumber\\&&
     -\Li\left((2+\sxi){1-w\over 4}\right)
     +\Li\left(\frac{2\sxi}{(2+\sxi)(1+w)}\right)
     -\Li\left(\frac{2\sxi}{(2+\sxi)(1-w)}\right)\nonumber\\&&
     +{2-\xi\over2}\ln\left(\frac{1+v}{1-v}\right)-v
\end{eqnarray}

\newpage

\section*{Appendix B}
\setcounter{equation}{0}
\def\theequation{B\arabic{equation}}
It is convenient to define the mass dependent variables $a:=2+\sqrt\xi$, 
$b:=2-\sqrt\xi$ and $w:=\sqrt{(1-\sqrt\xi)/(1+\sqrt\xi)}$. The rate 
functions $t_1,\ldots,t_{12}$ appearing in Eqs.~(\ref{eqn18}--\ref{eqn20}) 
are then given by
\begin{eqnarray}
t_1&:=&\ln\left(\frac{2\xi\sqrt\xi}{b^2(1+\sqrt\xi)}\right),\quad
t_2\ :=\ \ln\left(\frac{2\sqrt\xi}{1+\sqrt\xi}\right)\quad
  \Rightarrow\quad t_1-t_2\ =\ \ln\left(\frac\xi{b^2}\right)\\
t_3&:=&\ln\left(\frac{1+v}{1-v}\right)\\
t_4&:=&\Li(w)-\Li(-w)+\Li(\frac abw)-\Li(-\frac abw)\\
t_5&:=&\frac12\ln\left(\frac{a\sqrt\xi}{4(1+\sqrt\xi)}\right)
  \ln\left(\frac{1+v}{1-v}\right)
  +\Li\left(\frac{2\sqrt\xi}{a(1+w)}\right)
  -\Li\left(\frac{2\sqrt\xi}{a(1-w)}\right)\,+\nonumber\\&&
  +\Li\left(\frac{1+w}2\right)-\Li\left(\frac{1-w}2\right)
  +\Li\left(\frac{a(1+w)}4\right)-\Li\left(\frac{a(1-w)}4\right)\\
t_6&:=&\ln^2(1+w)+\ln^2(1-w)+\ln\left(\frac a8\right)\ln(1-w^2)
  \,+\nonumber\\&&
  +\Li\left(\frac{2\sqrt\xi}{a(1+w)}\right)
  +\Li\left(\frac{2\sqrt\xi}{a(1-w)}\right)
  -2\Li\left(\frac{2\sqrt\xi}a\right)\,+\nonumber\\&&
  +\Li\left(\frac{1+w}2\right)+\Li\left(\frac{1-w}2\right)
  -2\Li\left(\frac12\right)\,+\nonumber\\&&
  +\Li\left(\frac{a(1+w)}4\right)+\Li\left(\frac{a(1-w)}4\right)
  -2\Li\left(\frac a4\right)\\
t_7&:=&2\ln\left(\frac{1-\xi}{2\xi}\right)\ln\left(\frac{1+v}{1-v}\right)
  -\Li\left(\frac{2v}{(1+v)^2}\right)
  +\Li\left(-\frac{2v}{(1-v)^2}\right)\,+\nonumber\\&&
  -\frac12\Li\left(-\left(\frac{1+v}{1-v}\right)^2\right)
  +\frac12\Li\left(-\left(\frac{1-v}{1+v}\right)^2\right)\,+\\&&
  +\Li\left(\frac{2w}{1+w}\right)-\Li\left(-\frac{2w}{1-w}\right)
  -2\Li\left(\frac{w}{1+w}\right)+2\Li\left(-\frac{w}{1-w}\right)
  \,+\nonumber\\&&
  +\Li\left(\frac{2aw}{b+aw}\right)-\Li\left(-\frac{2aw}{b-aw}\right)
  -2\Li\left(\frac{aw}{b+aw}\right)+2\Li\left(-\frac{aw}{b-aw}\right)
  \nonumber\\
t_8&:=&\ln\left(\frac\xi 4\right)\ln\left(\frac{1+v}{1-v}\right)
  +\Li\left(\frac{2v}{1+v}\right)-\Li\left(-\frac{2v}{1-v}\right)-\pi^2\\
t_9&:=&2\ln\left(\frac{2(1-\xi)}{\sqrt\xi}\right)
  \ln\left(\frac{1+v}{1-v}\right)
  +2\left(\Li\left(\frac{1+v}2\right)-\Li\left(\frac{1-v}2\right)\right)
  \,+\nonumber\\&&
  +3\left(\Li\left(-\frac{2v}{1-v}\right)
  -\Li\left(\frac{2v}{1+v}\right)\right)\\
t_{10}&:=&\ln\left(\frac4\xi\right),\quad
t_{11}\ :=\ \ln\left(\frac{4(1-\sqrt\xi)^2}\xi\right),\quad
t_{12}\ :=\ \ln\left(\frac{4(1-\xi)}\xi\right)
\end{eqnarray}

\section*{Appendix C}
\setcounter{equation}{0}
\def\theequation{C\arabic{equation}}
In this Appendix we generalize the cross-section expressions in the main 
text to include transverse and longitudinal beam polarization 
effects. The cross section can be written in a matrix factorized form 
as~\cite{forward27}
\begin{equation}\label{eqnc1}
d\sigma=\left(\frac{2\pi\alpha}{q^2}\right)^2\sum_{i,j=1}^4g_{ij}(q^2)
  L^i\cdot H^j dPS,
\end{equation}
where $\alpha=1/137$ is the fine structure constant, $q^2=4E_{\rm beam}^2$ 
the total quared c.m.~beam energy, $dPS$ the Lorentz-invariant phase space 
element and $L\cdot H$ stands for $L_{\mu\nu}H^{\mu\nu}$. The indices $i$ 
and~$j$ ($i,j=1,2,3,4$) run over the four independent components of the 
hadron tensor $H^i_{\mu\nu}$ (defined in Eq.~(\ref{eqn2})) and the lepton 
tensor $L^i_{\mu\nu}$, resp.\ (defined accordingly). $H^i_{\mu\nu}$ stands 
for the final hadron tensor regardless of whether one is considering polarized 
or unpolarized final states. In the case treated here (polarized quark), 
the hadron tensor would carry an additional index labelling the three 
independent polarization components of the quark, i.e.\ in the notation of 
the mean part of the paper one would have 
$H_{\mu\nu}^i\rightarrow H_{\mu\nu}^{i(\ell)}$ etc. However, since the 
discussion in this Appendix is quite general, we drop the explicit 
reference to possible polarization dependences in the hadron tensor.

The coefficients of the energy dependent electro-weak coupling matrix 
$g_{ij}(q^2)$ are given below. The factorized form of Eq.~(\ref{eqnc1}) has 
the advantage that the various parts determining the cross section are 
separated into three modular pieces: $L^i_{\mu\nu}$ contains the angular 
dependences and the beam polarization parameters, $H^i_{\mu\nu}$ contains 
the hadron dynamics of the final state and $g_{ij}(q^2)$ comprises the model 
dependent parameters of the underlying electro-weak theory.

The electro-weak coupling matrix $g_{ij}(q^2)$ entering in Eq.~(\ref{eqnc1}) 
is given by
\begin{eqnarray}
g_{11}&=&Q_f^2-2Q_fv_ev_f\real\cz+(v_e^2+a_e^2)(v_f^2+a_f^2)|\cz|^2,\nonumber\\
g_{12}&=&Q_f^2-2Q_fv_ev_f\real\cz+(v_e^2+a_e^2)(v_f^2-a_f^2)|\cz|^2,\nonumber\\
g_{13}&=&-2Q_fv_ea_f\imag\cz,\\
g_{14}&=&2Q_fv_ea_f\real\cz-2(v_e^2+a_e^2)v_fa_f|\cz|^2,\nonumber
\end{eqnarray}

\newpage

\begin{eqnarray}
g_{21}&=&q_f^2-2Q_fv_ev_f\real\cz+(v_e^2-a_e^2)(v_f^2+a_f^2)|\cz|^2,\nonumber\\
g_{22}&=&q_f^2-2Q_fv_ev_f\real\cz+(v_e^2-a_e^2)(v_f^2-a_f^2)|\cz|^2,\nonumber\\
g_{23}&=&-2Q_fv_ea_f\imag\cz,\nonumber\\
g_{24}&=&2Q_fv_ea_f\real\cz-2(v_e^2-a_e^2)v_fa_f|\cz|^2,\nonumber\\
\nonumber\\
g_{31}&=&-2Q_fa_ev_f\imag\cz,\nonumber\\
g_{32}&=&-2Q_fa_ev_f\imag\cz,\nonumber\\
g_{33}&=&2Q_fa_ea_f\real\cz,\addtocounter{equation}{-1}\\
g_{34}&=&2Q_fa_ea_f\imag\cz,\nonumber\\
\nonumber\\
g_{41}&=&2Q_fa_ev_f\real\cz-2v_ea_e(v_f^2+a_f^2)|\cz|^2,\nonumber\\
g_{42}&=&2Q_fa_ev_f\real\cz-2v_ea_e(v_f^2-a_f^2)|\cz|^2,\nonumber\\
g_{43}&=&2Q_fa_ea_f\imag\cz,\nonumber\\
g_{44}&=&-2Q_fa_ea_f\real\cz+4v_ea_ev_fa_f|\cz|^2\nonumber
\end{eqnarray}
where $\cz(q^2)=gM_Z^2q^2/(q^2-M_Z^2+iM_Z\Gamma_Z)^{-1}$, with $M_Z$ and 
$\Gamma_Z$ the mass and width of the $Z^0$ and
$g=G_F(8\sqrt 2\pi\alpha)^{-1}\approx 4.49\cdot 10^{-5}\mbox{\rm GeV}^{-2}$.
$Q_f$ are the charges of the final state quarks to which the electro-weak 
currents directly couple; $v_e$ and $a_e$, $v_f$ and $a_f$ are the 
electro-weak vector and axial vector coupling constants. For example, in 
the Weinberg-Salam model, one has $v_e=-1+4\sin^2\theta_W$, $a_e=-1$ for 
leptons, $v_f=1-\frac83\sin^2\theta_W$, $a_f=1$ for up-type quarks 
($Q_f=\frac23$), and $v_f=-1+\frac43\sin^2\theta_W$, $a_f=-1$ for down-type 
quarks ($Q_f=-\frac13$). The left- and right-handed coupling constants are 
then given by $g_L=v+a$ and $g_R=v-a$, respectively. In the purely 
electromagnetic case one has $g_{11}=g_{12}=g_{21}=g_{22}=Q_f^2$ and all 
other $g_{r'r}=0$. The terms linear in $\real\cz$ and $\imag\cz$ come from 
$\gamma-Z^0$ interference, whereas the terms proportional to $|\cz|^2$ 
originate from $Z$-exchange.

\subsection*{C.1\quad Lepton tensor}
We first specify the lepton tensor in the laboratory frame denoted by a 
prime. The nonvanishing elements of the lepton tensor $L'^i_{\mu\nu}$ 
(with $i=1,2,3,4$ defined in analogy to Eq.~(\ref{eqn2})) are given by
\begin{eqnarray}
L'^1_{11}\ =\ L'^1_{22}\ =\ (1-h^-h^+),&&\mbox{\qquad\qquad s}\nonumber\\
L'^1_{12}\ =\ -L'^1_{21}\ =\ -i(h^--h^+),&&\mbox{\qquad\qquad a}\nonumber\\
L'^2_{11}\ =\ L'^2_{22}\ =\ -(\xi^-_{x'}\xi^+_{x'}-\xi^-_{y'}\xi^+_{y'}),
  &&\mbox{\qquad\qquad s}\nonumber\\
L'^2_{12}\ =\ L'^2_{21}\ =\ -(\xi^-_{x'}\xi^+_{y'}+\xi^-_{y'}\xi^+_{x'}),
  &&\mbox{\qquad\qquad s}\label{eqnc2}\\
L'^3_{11}\ =\ -L'^3_{22}\ =\ (\xi^-_{x'}\xi^+_{y'}+\xi^-_{y'}\xi^+_{x'}),
  &&\mbox{\qquad\qquad s}\nonumber\\
L'^3_{12}\ =\ L'^3_{21}\ =\ -(\xi^-_{x'}\xi^+_{x'}-\xi^-_{y'}\xi^+_{y'}),
  &&\mbox{\qquad\qquad s}\nonumber\\
L'^4_{11}\ =\ L'^4_{22}\ =\ (h^--h^+),&&\mbox{\qquad\qquad s}\nonumber\\
L'^4_{12}\ =\ -L'^4_{21}\ =\ -i(1-h^-h^+).&&\mbox{\qquad\qquad a}\nonumber
\end{eqnarray}
Here we have indicated the $\mu\leftrightarrow\nu$ symmetry properties 
(s=symmetric, a=antisymmetric) of the various components of $L'_{\mu\nu}$ 
which will be of later convenience when considering the contraction with 
the hadronic tensor. The $\xi^\pm_{x'}$ and $\xi^\pm_{y'}$ denote the 
transverse beam polarization in and out of the accelerator plane, and 
$\xi^\pm_{z'}$($=\mp h^\pm$) the longitudinal polarization (helicity) of 
$e^\pm$. For natural transverse beam polarization of degree $P$ in the case 
of a circular accelerator one has $\vec\xi^\pm=(0,\pm P,0)$.

\begin{table}\begin{center}
\caption{\label{tab1}Independent helicity components of the hadron tensor
$H_{\mu\nu}$ in the spherical basis (second column) and in the Cartesian basis
(third column). Column 4 gives the $\mu\leftrightarrow\nu$ symmetry of the
nine components and column 5 lists the parity of the products of contributing
currents ($pc$: $VV,AA$ and $pv$: $V\!A,AV$) for the unpolarized case and in
round brackets for the polarized case.}

\vspace{12pt}
\begin{tabular}{|l||l|l|c|r|}
\hline
&spherical components&Cartesian components&&\\\hline\hline
$H_U$&$H_{++}+H_{--}$&$H_{11}+H_{22}$&s&$pc$($pv$)\\\hline
$H_L$&$H_{00}$&$H_{33}$&s&$pc$($pv$)\\\hline
$H_T$&$\frac12(H_{+-}+H_{-+})$&$\frac12(-H_{11}+H_{22})$
  &s&$pc$($pv$)\\\hline
$H_I$&$\frac14(H_{+0}+H_{0+}-H_{-0}-H_{0-})$
  &$\frac{-1}{2\sqrt 2}(H_{31}+H_{13})$&s&$pc$($pv$)\\\hline
$H_9$&$-\frac i4(H_{+0}-H_{0+}-H_{-0}+H_{0-})$
  &$\frac{-i}{2\sqrt 2}(H_{31}-H_{13})$&a&$pc$($pv$)\\\hline
$H_F$&$H_{++}-H_{--}$&$-i(H_{12}-H_{21})$&a&$pv$($pc$)\\\hline
$H_A$&$\frac14(H_{+0}+H_{0+}+H_{-0}+H_{0-})$
  &$\frac{-i}{2\sqrt 2}(H_{23}-H_{32})$&a&$pv$($pc$)\\\hline
$H_4$&$-\frac i2(H_{+-}-H_{-+})$&$-\frac12(H_{12}+H_{21})$
  &s&$pv$($pc$)\\\hline
$H_5$&$-\frac i4(H_{+0}-H_{0+}+H_{-0}-H_{0-})$
  &$\frac{-1}{2\sqrt 2}(H_{23}+H_{32})$&s&$pv$($pc$)\\\hline
\end{tabular}
\end{center}\end{table}

\subsection*{C.2\quad Hadron tensor}
In the overall c.m.~system that we are considering, only the three space 
components of the hadron tensors $H^i_{mn}$ ($m,n=1,2,3$) contribute to 
the annihilation cross section in Eq.~(\ref{eqnc1}). Also, since the hadron 
tensors $H^i_{mn}$ are hermitean, there are in general nine real 
independent components $H^i_{mn}$ for each $i=1,2,3,4$. We introduce six 
symmetric and three antisymmetric real combinations under $m\leftrightarrow n$ 
(dropping the superscript $i$ for compactness) in Table~\ref{tab1}, with 
$H_{\sigma\sigma'}=\eps^\mu(\sigma)H_{\mu\nu}\eps^{*\nu}(\sigma')$, where
\begin{equation}
\eps^\mu(\pm)=\frac1{\sqrt2}(0;\mp 1,-i,0),\qquad \eps^\mu(0)=(0;0,0,1).
\end{equation}
Table~\ref{tab1} contains the sperical as well as the Cartesian components of
the nine independent hadron tensor components in the c.m.\ system. We also
list the $\mu\leftrightarrow\nu$ symmetry properties of the components as well 
as the parity nature of the product of currents in the helicity system 
($z$-axis in event plane!).

\begin{figure}\begin{center}
\epsfig{figure=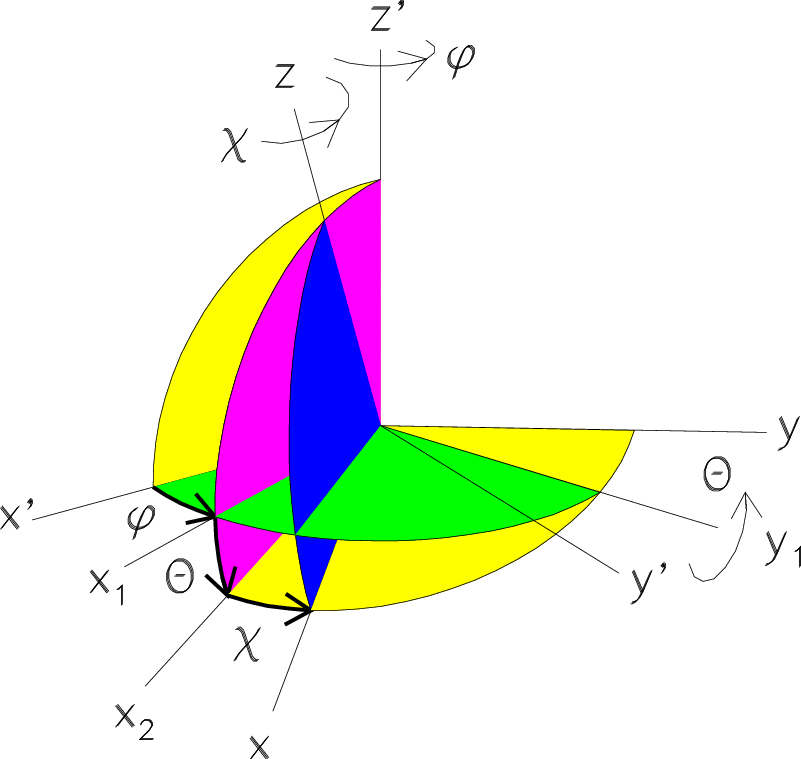, scale=0.8}
\caption{\label{fig4}Definition of the Euler angles $\varphi$, $\theta$ and
$\chi$ relating the beam frame ${\cal O}_{x'y'z'}$ and the event frame
${\cal O}_{xyz}$. The three successive Euler rotations from the beam frame to
the event frame are $R_{z'}(\varphi)$ followed by $R_{y_1}(\theta)$ and then
by $R_z(\chi)$}
\end{center}\end{figure}
\subsection*{C.3\quad Angular correlations}
We are now in the position to consider the beam-event correlations that 
arise from the contraction of the lepton and hadron tensors. First we have 
to rotate the lepton tensors $L'_{mn}$ given in Eq.~(\ref{eqnc2}) to the 
event plane. This is achieved by the Euler rotation illustrated in
Fig.~\ref{fig4},
\begin{equation}
L_{kl}(\varphi,\theta,\chi)=R_{km}(\varphi,\theta,\chi)
  R_{ln}(\varphi,\theta,\chi)L'_{mn},
\end{equation}
where $\varphi$, $\theta$ and $\chi$ are the Euler angles describing the 
rotation from the beam-system to the event-system, and the Euler rotation 
matrix is given by
\medskip
\begin{eqnarray}
\lefteqn{R(\varphi,\theta,\chi)}\\[12pt]
&=&\pmatrix{\cos\varphi\cos\theta\cos\chi-\sin\varphi\sin\chi&
  \sin\varphi\cos\theta\cos\chi+\cos\varphi\sin\chi&-\sin\theta\cos\chi\cr
  -\cos\varphi\cos\theta\sin\chi-\sin\varphi\cos\chi&
  -\sin\varphi\cos\theta\sin\chi+\cos\varphi\cos\chi&\sin\theta\sin\chi\cr
  \cos\varphi\sin\theta&\sin\varphi\sin\theta&\cos\theta\cr}.\nonumber
\end{eqnarray}
\medskip\\
Let us note, though, that the contraction $L_{mn}(\varphi,\theta,\chi)H_{mn}$ 
is simplified by separately considering the symmetric and antisymmetric 
parts of $L_{mn}$ and $H_{mn}$ according to the classification given in 
Table~\ref{tab1}. Such symmetry considerations are also very helpful when one
wants to assess quickly the influence of beam polarization on the measurement
of the various hadron tensor components. One obtains (omitting again the 
superscript $i$ on the hadron tensor):
\begin{eqnarray}
\frac34L^1_{mn}H_{mn}&=&(1-h^-h^+)H_A+(h^--h^+)H_D,\nonumber\\
\frac34L^2_{mn}H_{mn}&=&X(\varphi)H_B+Y(\varphi)H_C,\nonumber\\
\frac34L^3_{mn}H_{mn}&=&X(\varphi)H_C-Y(\varphi)H_B,\label{eqnc3}\\
\frac34L^4_{mn}H_{mn}&=&(1-h^-h^+)H_D+(h^--H^+)H_A,\nonumber
\end{eqnarray}
where
\begin{eqnarray}
H_A&=&\frac38(1+\cos^2\theta)H_U+\frac34\sin^2\theta H_L
  +\frac34\sin^2\theta\cos 2\chi H_T\nonumber\\&&
  -\frac34\sin^2\theta\sin 2\chi H_4
  -\frac3{2\sqrt2}\sin 2\theta\cos\chi H_I
  +\frac3{2\sqrt2}\sin 2\theta\sin\chi H_5,\nonumber\\
H_B&=&\frac38\sin^2\theta H_U-\frac34\sin^2\theta H_L
  +\frac34(1+\cos^2\theta)\cos 2\chi H_T\nonumber\\&&
  -\frac34(1+\cos^2\theta)\sin 2\chi H_4
  +\frac3{2\sqrt2}\sin 2\theta\cos\chi H_I
  -\frac3{2\sqrt2}\sin 2\theta\sin\chi H_5,\nonumber\\
H_C&=&\frac32\cos\theta\sin 2\chi H_T+\frac32\cos\theta\cos 2\chi H_4
  \label{eqnc4}\\&&
  +\frac3{\sqrt2}\sin\theta\sin\chi H_I
  +\frac3{\sqrt2}\sin\theta\cos\chi H_5,\nonumber\\
H_D&=&\frac34\cos\theta H_F-\frac3{\sqrt2}\sin\theta\cos\chi H_A
  +\frac3{\sqrt2}\sin\theta\sin\chi H_9,\nonumber
\end{eqnarray}
with all hadron tensors referring to the hadronic event system and
\begin{eqnarray}
X(\varphi)&=&(\xi^-_{x'}\xi^+_{x'}-\xi^-_{y'}\xi^+_{y'})\cos 2\varphi
  +(\xi^-_{x'}\xi^+_{y'}+\xi^-_{y'}\xi^+_{x'})\sin 2\varphi,\nonumber\\
Y(\varphi)&=&-(\xi^-_{x'}\xi^+_{x'}-\xi^-_{y'}\xi^+_{y'})\sin 2\varphi
  +(\xi^-_{x'}\xi^+_{y'}+\xi^-_{y'}\xi^+_{x'})\cos 2\varphi.
\end{eqnarray}
Note that $H_B$ and $H_C$ contain no more hadronic structure than $H_A$. 
This means that the presence of transverse beam polarization does not 
increase the number of possible hadronic measurements. On the other hand, 
transverse beam polarization is helpful in disentangling the hadronic 
structure.

Eq.~(\ref{eqnc3}) gives the most general structure of possible beam-event 
correlations in the presence of beam polarization and final state 
polarization effects. The whole angular dependence on the relative 
beam-event angles $\varphi$, $\theta$ and $\chi$ is explicitly exhibited in 
Eqs.~(\ref{eqnc3}) and~(\ref{eqnc4}), since the hadronic helicity structure 
functions $H_\alpha$ are evaluated in the event frame and thus are 
independent of these angles. We repeat that, when we have been referring 
to the hadron tensor in this Appendix, this means either the unpolarized 
hadron tensor or the polarized hadron tensor carrying any number of 
additional polarization indices (in our case the quark's polarization 
index $m$ which can take the three values $m=\ \perp$ (transversal 
perpendicular), $N$ (transverse normal) and $\ell$ (longitudinal)). Also, in 
this paper we have only considered the $(U)$-, $(L)$- and $(F)$-components 
of the hadron tensor in Eq.~(\ref{eqnc4}) which remain after azimuthal 
$\varphi$- and $\chi$-averaging.


\begin{thebibliography}{99}
\bibitem{forward1}J.G.~K\"orner, A.~Pilaftsis and M.M.~Tung,
  Z.~Phys. {\bf C63} (1994) 575
\bibitem{forward2}M.M.~Tung, Phys.~Rev.\ {\bf D52} (1995) 1353
\bibitem{forward3}S.~Groote, J.G.~K\"orner and M.M.~Tung,
  Z.~Phys.\ {\bf C70} (1996) 281
\bibitem{forward4}S.~Groote and  J.G.~K\"orner,\\
  ``Transverse Polarization of Top Quarks Produced in $e^+e^-$-Annihilation 
  at $O(\as)$'',\\
  Mainz preprint MZ-TH/95-17, hep-ph/9508399, to be published in Z.~Phys.~C.
\bibitem{forward5}G.~Grunberg, Y.J.~Ng and S.H.H.~Tye,
  Phys.~Rev.\ {\bf D21} (1980) 62
\bibitem{forward6}J.~Schwinger, {\it Particles, Sources and Fields\/}\\
  (Addison Wesley, New York, 1973), Vol.~II, Sec.~5.4
\bibitem{forward7}J.~Jerz\'ak, E.~Laermann and P.M.~Zerwas,\\
  Phys.~Rev.\ {\bf D25} (1982) 1218; Erratum: {\bf D36} (1987) 310(E)
\bibitem{forward8}F.~Abe et al. (CDF Collaboration),
  Phys.~Rev.~Lett.\ {\bf 75} (1995) 3997
\bibitem{forward9}R.~Harlander, M.~Je\.zabek, J.H.~K\"uhn, T.~Teubner,
  Phys.~Lett.\ {\bf B346} (1995) 137;\\
  M.~Je\.zabek, Karlsruhe preprint TTP 95-12, hep-ph/9503461;\\
  M.~Je\.zabek, R.Harlander, J.H.~K\"uhn, M.~Peter,\\
  Karlsruhe preprint TTP 95-47 (1995), hep-ph/9512409
\bibitem{forward10}J.B.~Stav and H.A.~Olsen, 
  Phys.~Rev.\ {\bf D54} (1996) 817; see also\\
  J.B.~Stav and H.A.~Olsen, Phys.~Rev.\ {\bf D52} (1995) 1359;\\
  J.B.~Stav and H.A.~Olsen, Z.~Phys.\ {\bf C57} (1993) 519
\bibitem{forward11}A.B.~Arbuzov, D.Y.~Bardin and A.~Leike,\\
  Mod.~Phys.~Lett.\ {\bf A7} (1992) 2029; Erratum: {\bf A9} (1994) 1515
\bibitem{forward12}M.B.~Voloshin, Int.~J.~of Mod.~Phys.\ 
  {\bf A10} (1995) 2865
\bibitem{forward13}G.~Rodrigo and A.~Santamaria, Phys.~Lett.\ 
  {\bf B313} (1993) 441;\\
  W.~Bernreuther, Aachen preprint PITHA-94/31, hep-ph/9409390
\bibitem{forward14}T.D.~Lee and M.~Nauenberg, 
  Phys.~Rev.\ {\bf B6} (1964) 1594;\\
  A.V.~Smilga, Commun.~Nucl.~Part.~Phys.\ {\bf 20} (1991) 69
\bibitem{forward15}B.~Falk and L.M.~Sehgal,
  Phys.~Lett.\ {\bf B325} (1994) 509
\bibitem{forward16}W.~Siegel, Phys.~Lett.\ {\bf B84} (1979) 193
\bibitem{forward17}J.G.~K\"orner, S.~Sakakibara and G.~Schuler,
  Phys.~Lett.\ {\bf B194} (1987) 125
\bibitem{forward18}J.G.~K\"orner and M.M.~Tung,
  Z.~Phys.\ {\bf C64} (1994) 255
\bibitem{forward19}J.G.~K\"orner, D.~Kreimer and K.~Schilcher,
  Z.~Phys.\ {\bf C54} (1992) 503
\bibitem{forward20}J.G.~K\"orner, G.~Schuler, G.~Kramer and B.~Lampe,
  Z.~Phys.\ {\bf C32} (1986) 181
\bibitem{forward21}S.~Jadach and Z.~Was,
  Acta~Phys.~Pol.\ {\bf B15} (1984) 1151;\\
  Z.~Was, Acta~Phys.~Pol.\ {\bf B18} (1987) 1099
\bibitem{forward22}J.H.~K\"uhn, Acta Physica Austriaca,
  Suppl.~XXIV (1982) 203;\\
  R.H.~Dalitz and G.R.~Goldstein, Phys.~Rev.\ {\bf D45} (1992) 1531;\\
  F.E.~Close, J.G.~K\"orner, R.J.N.~Phillips and
  D.J.~Summers,\\ J.~Phys.\ {\bf G18} (1992) 1716;\\
  T.~Mannel and G.A.~Schuler, Phys.~Lett.\ {\bf B279} (1992) 194;\\
  B.~Mele and G.~Altarelli, Phys.~Lett.\ {\bf B299} (1993) 345;\\
  B.~Mele, Mod.~Phys.~Lett.\ {\bf A9} (1994) 1239;\\
  G.~Bonvicini and L.~Randall, Phys.~Rev.~Lett.\ {\bf 73} (1994) 392;\\
  A.~Czarnecki, M.~Je\.zabek, J.G.~K\"orner and J.H.~K\"uhn,\\
  Phys.~Rev.~Lett.\ {\bf 73} (1994) 384;\\
  A.~Czarnecki and M.~Je\.zabek, Nucl.~Phys.\ {\bf B427} (1994) 3;\\
  J.K.~Kim and Y.G.~Kim, Phys.~Rev.\ {\bf D52} (1995) 5352;\\
  C.~Diaconu, J.G.~K\"orner, D.~Pirjol and M.~Talby,
  Phys.~Rev.\ {\bf D53} (1996) 6186;\\
  C.~Diaconu, J.G.~K\"orner, D.~Pirjol and M.~Talby,
  ``Spin-Momentum Correlations in Inclusive Semileptonic Decays of 
  Polarized $\Lambda_b$-Baryons'', to be published
\bibitem{forward23}M.~Anselmino, P.~Kroll and B.~Pire, 
  Phys.~Lett.\ {\bf B167} (1986) 113;\\
  J.H.~K\"uhn, A.~Reiter and P.M.~Zerwas,
  Nucl.~Phys.\ {\bf B272} (1986) 560
\bibitem{forward24}W.~Bernreuther, J.P.~Ma and T.~Schr\"oder,
  Phys.~Lett.\ {\bf B297} (1992) 318;\\
  W.~Bernreuther, O.~Nachtmann, P.~Overmann and T.~Schr\"oder,\\
  Nucl.~Phys.\ {\bf B388} (1992) 53; Erratum: {\bf B406} (1993) 516;\\
  A.F.~Falk and M.E.~Peskin, Phys.~Rev.\ {\bf D49} (1994) 3320;\\
  J.G.~K\"orner and M.~Kr\"amer, Phys. Lett.\ {\bf B275} (1992) 495;\\
  P.~Bialas, J.G.~K\"orner, M.~Kr\"amer and K.~Zalewski,
  Z.~Phys.\ {\bf C57} (1993) 115
\bibitem{forward25}A.~Bartl, E.~Christova and W.~Majerotto,\\
  University of Wien preprint UWThPh-1995-9
\bibitem{forward26}C.R.~Schmidt, ``Top Quark Production and Decay at 
  Next-to-leading Order in $e^+e^-$ Annihilation'' (1995)
  hep-ph/9504434, to be published in Phys.~Rev.~D
\bibitem{forward27}J.G.~K\"orner and D.H.~Schiller,
  preprint DESY-81-043 (1981)
\end{thebibliography}
\end{document}